\documentclass[preprint]{sigplanconf}

\usepackage{amsmath, amsfonts, amssymb}
\usepackage{color}
\usepackage{graphicx}
\usepackage{amsthm}

\usepackage{url}
\usepackage{xspace}
\usepackage{picins}
\usepackage{multicol}

\newcommand{\Reals}{\ensuremath{\mathbb{R}}}

\newcommand{\Nat}{\ensuremath{\mathbb{N}}}

\newcommand{\Prob}[1]{\ensuremath{\mathbb{P}[#1]}}

\newcommand{\Esp}[1]{\ensuremath{\mathbb{E}[#1]}}

\newcommand{\Distrib}[1]{\ensuremath{f_{#1}}}

\newcommand{\cumulative}[1]{\ensuremath{\mathcal{F}_{#1}}}

\newcommand{\Zn}{\ensuremath{Z^{(n)}}}

\newcommand{\Gain}[1]{\ensuremath{\mathcal{G}_{#1}}}

\definecolor{purple}{rgb}{0.5,0,0.5}

\newcommand{\eg}{\emph{e.g.,}\xspace}
\newcommand{\ie}{\emph{i.e.,}\xspace}

\newcommand{\ai}{\textsc{All-Interval}\xspace}
\newcommand{\ms}{\textsc{Magic-Square}\xspace}
\newcommand{\ca}{\textsc{Costas Array}\xspace}

\def\CF#1{\multicolumn{1}{||c|}{#1}}

\def\C#1{\multicolumn{1}{c|}{#1}}
\def\D#1{\multicolumn{1}{c||}{#1}}
\def\N#1{\multicolumn{1}{@{~}c@{~}}{#1}}
\def\NN#1{\multicolumn{1}{@{~}c@{~}}{#1}}

\newtheorem{definition}{Definition}

\begin{document}


\titlebanner{\ }        
\preprintfooter{Parallel Las Vegas Algorithms}   

\title{Prediction of Parallel Speed-ups for Las Vegas Algorithms}


\authorinfo{Charlotte Truchet}
           {LINA, UMR 6241 / University of Nantes}
           {Charlotte.Truchet@univ-nantes.fr}
\authorinfo{Florian Richoux}
           {LINA, UMR 6241 / University of Nantes\\JFLI, CNRS / University of Tokyo}
           {florian.richoux@univ-nantes.fr}
\authorinfo{Philippe Codognet}
           {JFLI, CNRS/UPMC/University of Tokyo}
           {codognet@is.s.u-tokyo.ac.jp}

\maketitle

\begin{abstract}

  We propose a probabilistic model  for the parallel execution of {\it
    Las  Vegas algorithms}, \ie randomized algorithms  whose runtime
  might vary from one execution  to another, even with the same input.
  This   model   aims   at   predicting  the   parallel   performances
  (\ie  speedups)  by  analysis   the  runtime  distribution  of  the
  sequential runs  of the algorithm.   Then, we study in  practice the
  case  of   a  particular  Las  Vegas   algorithm  for  combinatorial
  optimization,  on  three classical  problems,  and  compare with  an
  actual parallel  implementation up to  256 cores.  We show  that the
  prediction can be quite  accurate, matching the actual speedups very
  well up to  100 parallel cores and then with  a deviation of about 20\%
  up to 256 cores.
\end{abstract}

\category{D}{1.3}{Parallel programming}
\category{F}{1.2}{Parallelism and concurrency}
\category{G}{1.6}{Stochastic programming}
\category{G}{3}{Probabilistic algorithms (including Monte Carlo)}
\category{G}{2.1}{Combinatorial algorithms}

\terms
Theory, Algorithms, Performance.


\section{Introduction}

We will consider in this  paper {\it Las Vegas algorithms}, introduced
a few decades ago  by~\cite{Babai79}, i.e. randomized algorithms whose
runtime might vary  from one execution to another,  even with the same
input. An  important class  of Las Vegas  algorithms is the  family of
{\it Stochastic Local  Search} methods~\cite{Hoos-book}.  They have
been  used  in  Combinatorial  Optimization  for  finding  optimal  or
near-optimal  solutions  for  several decades  \cite{aarts:1997:lsco},
stemming from  the pioneering  work of Lin  on the  Traveling Salesman
Problem  \cite{Lin65}.   Theses methods are  now  widely  used in  combinatorial
optimization to solve real-life problems  when the search space is too
large  to be  explored by  complete  search algorithm,  such as  Mixed
Integer Programming or Constraint Solving, c.f. \cite{handbook-approx}.

In  the last  years, several  proposal for  implementing  local search
algorithms on  parallel computer have been proposed,  the most popular
being  to  run  several  competing  instances  of  the  algorithms  on
different cores, with different  initial conditions or parameters, and
let the  fastest process win over  others.  We thus  have an algorithm
with the minimal execution time among the launched processes.  This
lead to so-called independent  multi-walk  algorithms in the local search
community~\cite{Aarts95}  and portfolio algorithms  in the SAT
community  (satisfiability of Boolean formula)~\cite{portfolio01}.
This parallelization scheme can of course be generalized to any Las Vegas
algorithm.

The goal  of this paper is  to study the parallel  performances of Las
Vegas  algorithms under  this  independent multi-walk  scheme, and  to
predict  the performances of  the parallel  execution from  the runtime
distribution  of the  sequential runs  of a  given algorithm.  We will
confront  these  predictions  with  actual  speedups  obtained  for  a
parallel implementation of a local  search algorithm and show that the
prediction  can be quite  accurate, matching  the actual  speedup very
well up to 100 parallel cores  and then with a deviation limited to
about 20\% up to 256 cores.

The  paper is  organized  as  follows. Section 2 is devoted to  present
the definition  of Las Vegas algorithms, their parallel multi-walk
execution scheme, and the main idea for predicting the parallel  speedups.
Section~\ref{probabilistic-model} will detail
the  probabilistic model of  Las Vegas  algorithms and  their parallel
execution scheme. Section~\ref{local-search}  will present the example
of  local  search  algorithms  for combinatorial  optimization,  while
Section~\ref{benchmarks}  will detail the  benchmark problems  and the
sequential performances. Then, Section~\ref{prediction} will apply the
general probabilistic model to  the benchmark results and thus predict
their parallel speedup, which will be compared to actual speedups of a
parallel implementation in Section~\ref{analysis}.  A short conclusion
and future work end will the paper.

\section{Preliminaries}

\subsection{Las Vegas Algorithms}

We borrow the following definition from~\cite{Hoos-book}, Chapter 4.

\begin{definition}[Las Vegas Algorithm]
  An  algorithm A for  a problem  class $\Pi$  is a  (generalized) Las
  Vegas algorithm if and only if it has the following properties:
\begin{enumerate}
\item  If for  a given  problem instance  $\pi \in  \Pi$,  algorithm A
  terminates  returning a  solution $s$,  $s$  is guaranteed  to be  a
  correct solution of $\pi$.
\item For any given instance $\pi  \in \Pi$, the run-time of A applied
  to $\pi$ is a random variable.
\end{enumerate}
\end{definition}

This is  a slight  generalization of the  classical definition,  as it
includes algorithms which are not guaranteed to return a solution.

A large class of Las Vegas  algorithms is the so-called family of {\it
  metaheuristics},  such as  Simulated Annealing,  Genetic Algorithms,
Tabu Search,  Swarm Optimization, Ant-Colony  optimization, etc, which
have been applied to different  sets of problems ranging from resource
allocation, scheduling, packing,  layout design, frequency allocation,
etc.

\subsection{Multi-walk Parallel Extension}
\label{multiwalk}

Parallel         implementation         of        local         search
metaheuristics~\cite{handbook-approx,metaheuristics}  has been studied
since the early 1990s, when parallel machines started to become widely
available~\cite{DBLP:conf/irregular/PardalosPMR95,Aarts95}.   With the
increasing availability of PC clusters in the early 2000s, this domain
became   active  again~\cite{JHeuristics2004,JHeuristics2002}.   Apart
from domain-decomposition methods and population-based method (such as
genetic algorithms),~\cite{Aarts95}  distinguishes between single-walk
and multi-walk methods for  Local Search.  Single-walk methods consist
in  using  parallelism  inside   a  single  search  process,  \eg  for
parallelizing   the   exploration  of   the   neighborhood  (see   for
instance~\cite{GPU2010} for such  a method making use of  GPUs for the
parallel   phase).    Multi-walk   methods  (parallel   execution   of
multi-start methods) consist  in developing concurrent explorations of
the  search space,  either  independently or  cooperatively with  some
communication between  concurrent processes. Sophisticated cooperative
strategies  for multi-walk methods  can be  devised by  using solution
pools~\cite{DBLP:journals/heuristics/CrainicGHM04},     but    require
shared-memory or emulation of  central memory in distributed clusters,
thus  impacting on  performance.  A  key  point is  that a  multi-walk
scheme  is easier to  implement on  parallel computers  without shared
memory   and    can   lead,   in   theory   at    least,   to   linear
speedups~\cite{Aarts95}.   However  this is  only  true under  certain
assumptions 
and we  will see  that we need  to develop  a more realistic  model in
order  to cope  with  the performance  actually  observed in  parallel
executions.

Let us now formally define a parallel multi-walk Las Vegas algorithm.

\begin{definition}[Multi-walk Las Vegas Algorithm]
  An algorithm A' for a problem class $\Pi$ is a (parallel) multi-walk
  Las Vegas algorithm if and only if it has the following properties:
\begin{enumerate}
\item It consists of $n$ instances of a sequential Las Vegas algorithm
  A for $\Pi$, say $A_1,...,A_n$.
\item If, for a given problem  instance $\pi \in \Pi$, there exists at
  least one $i \in [1,n]$ such that $A_i$ terminates, then let $A_{m},
  m  \in [1,n]$, be  the instance  of A  terminating with  the minimal
  runtime  and let  $s$ be  the  solution returned  by $A_{m}$.   Then
  algorithm  A' terminates  in the  same time  as $A_{m}$  and returns
  solution $s$.
\item If, for a given problem  instance $\pi \in \Pi$, all $A_i, i \in
  [1,n]$, do not terminate then A' does not terminate.
\end{enumerate}
\end{definition}

\subsection{How to Estimate Parallel Speedup ?}

The multi-walk  parallel scheme is  rather simple, yet it  provides an
interesting test-case  to study how Las Vegas  algorithms can scale-up
in parallel. Indeed runtime will  vary among the processes launched in
parallel and  the overall  runtime will be  that of the  instance with
minimal  execution time  (i.e.  "long" runs  are  killed by  "shorter"
ones).   The question  is thus  to quantify  the (relative)  notion of
short and  long runs and  their probability distribution.   This might
gives us a key to quantify the expected parallel speed-up.  Obviously,
this can  be observed from the sequential behavior of the  algorithm, and
more  precisely from  the proportion  of long  and short  runs  in the
sequential runtime distribution.

In the  following, we  propose a probabilistic  model to  quantify the
expected speed-up of multi-walk  Las Vegas algorithms.  This makes it possible
to  give  a  general  formula  for  the  speed-up,  depending  on  the
sequential behavior  of the algorithm.   Our model is related  to {\it
  order      statistics},     a      rather     new      domain     of
statistics~\cite{david2003order},  which is  the statistics  of sorted
random draws.   Indeed, explicit formulas have been  given for several
well-known  distributions.    Relying  on  an   approximation  of  the
sequential  distribution,  we compute  the  average  speed-up for  the
multi-walk extension.   Experiments show that the  prediction is quite
good and  opens the  way for defining  more accurate models  and apply
them to larger classes of algorithms.

Previous  works  \cite{Aarts95}  studied  the  case  of  a  particular
distribution   for   the   sequential   algorithm,   the   exponential
distribution. This case is ideal and the best possible, as it yields a
linear speed-up.  Our model makes it possible to approximate Las Vegas algorithms
by other types of distribution, such as a  shifted  exponential distribution
or a lognormal distribution.
In the last two cases the speed-up is no longer linear, but admits a finite limit
when the number of processors tends toward infinity.
We will see that it fits experimental data for some problems.

\section{Probabilistic Model}
\label{probabilistic-model}

Local Search algorithms are stochastic processes. They include several
random  components: choice of  an initial  configuration, choice  of a
move among several candidates, plateau mechanism, random restart, etc.
In the  following, we will  consider the \emph{computation time}  of
an algorithm (whatever  it is) as  a random variable,  and use
elements  of  probability  theory  to study  its  multi-walk  parallel
version.  Notice that the computation time  is not  necessarily the cpu-time;
it  can also be the number  of iterations performed  during the execution
of the algorithm.

\subsection{\label{Min-Distribution} Min Distribution}

Consider a  given algorithm on a  given problem of a  given size, say,
the  \ms  $10 \times  10$.  Depending on  the  result  of some  random
components  inside the  algorithm, it  may find  a solution  after $0$
iterations,  $10$ iterations,  or  $10^6$ iterations.   The number  of
iterations of  the algorithm is  thus a discrete random  variable, let's call it
$Y$, with  values in \Nat. $Y$  can be studied  through its cumulative
distribution, which  is by definition, the  function \cumulative{Y} s.t.
$\cumulative{Y}(x)=\Prob{Y   \leq  x}$,   or  by its  distribution,
which   is   by definition   the   derivative     of     \cumulative{Y}:
$\Distrib{Y}=\cumulative{Y}'$.

It is often  more convenient to consider distributions  with values in
\Reals~because it makes calculations easier. For the same reason,
although $\Distrib{Y}$  is defined  in \Nat, we  will use  its natural
extension  to \Reals.   The  expectation of  the  computation is  then
defined as $\Esp{Y}=\int_0^\infty t \Distrib{Y}(t) dt$

\begin{figure}[t]
\begin{center}
\includegraphics[width=0.4\textwidth]{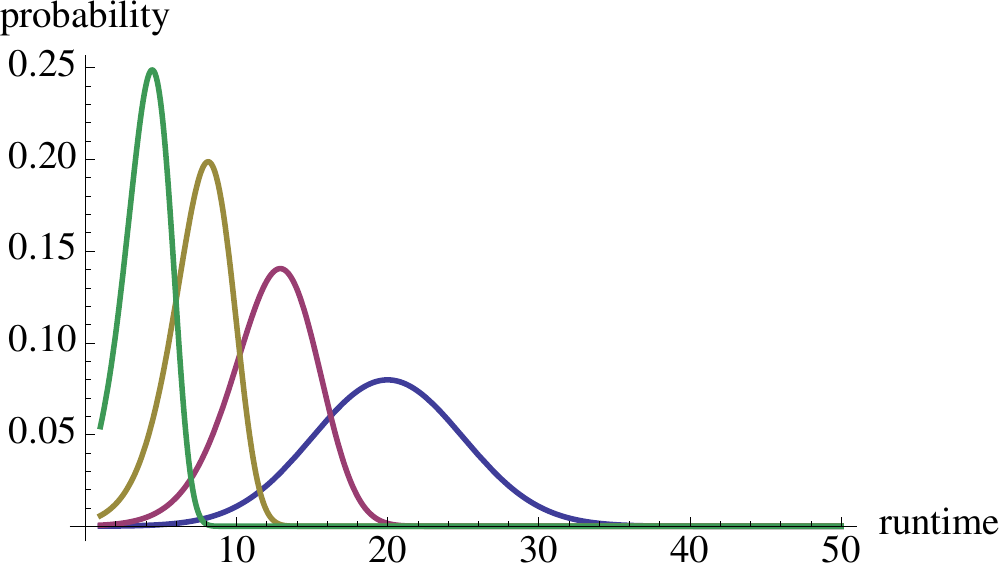}
\caption{\label{min-distribution}Distribution  of  \Zn,  in  the  case
  where $Y$ admits a  gaussian distribution (cut on $\mathbb{R}^-$ and
  renormalized). The blue curve is $Y$. The distributions of $\Zn$ are
  in  pink  for  $n=10$,  in  yellow  for $n=100$  and  in  green  for
  $n=1000$. }
\end{center}
\end{figure}

Assume that the base algorithm  is concurrently run in parallel on $n$
cores. In other words, over each core the running process is a fork of
the algorithm.  The  first process which finds a solution then kills all others.
and the algorithm terminates.
The $i$-th  process corresponds to a draw  of a random variable
$X_i$,  following  distribution  $\Distrib{Y}$.  The variables $X_i$  are  thus
independently  and identically  distributed (i.i.d.).  The computation
time of the whole parallel process is also a random variable, let \Zn,
with  a distribution  \Distrib{\Zn} that  depends both  on $n$  and on
\Distrib{Y}.  Since   all  the   $X_i$  are  i.i.d.,   the  cumulative
distribution $\cumulative{\Zn}$ can be computed as follows:
\begin{eqnarray*}
\cumulative{\Zn} & = & \Prob{\Zn \leq x} \\
 & = & \Prob{\exists i \in \{1...n\}, X_i \leq x}\\
& = & 1 - \Prob{\forall i \in \{1...n\}, X_i > x}\\
& = & 1 - \prod_{i=1}^{n} \Prob{X_i > x}\\
& = & 1 - \left( 1 - \cumulative{Y}(x)\right)^n \\
\end{eqnarray*}
which leads to:
\begin{eqnarray*}
\Distrib{\Zn}&=&\left(1 -( 1 - \cumulative{Y}\right)^n) ' \\
&=&n\Distrib{Y} (1-\cumulative{Y})^{n-1}\\
\end{eqnarray*}
Thus, knowing  the distribution  for the base  algorithm $Y$,  one can
calculate the distribution for $\Zn$. In the general case, the formula
shows that  the parallel algorithm  favors short runs, by  killing the
slower processes. Thus, we can expect that the distribution of $\Zn$ moves
toward  the  origin,  and  is  more  peaked.  As  an  example,  Figure
\ref{min-distribution} shows  this phenomenon when  the base algorithm
admits a gaussian distribution.

\subsection{Expectation and Speed-up}
\label{expectation-and-speedup}

The  model described  above gives  the probability  distribution  of a
parallelized version of any random algorithm. We can now calculate the
expectation for the parallel process with the following relation:

\begin{eqnarray*}
\Esp{\Zn} &=& \int_0^\infty t \Distrib{\Zn}(t) dt\\
&=& n \int_0^\infty t \Distrib{Y}(t) (1-\cumulative{Y}(t))^{n-1}dt \\
\end{eqnarray*}

Unfortunately, this does not lead  to a general formula for \Esp{\Zn}.
In   the  following,  we   will  study   it  for   different  specific
distributions.

To measure  the gain  obtained by parallelizing  the algorithm  on $n$
core, we will study the speed-up $\Gain{n}$ defined as:
$$\Gain{n}=\Esp{Y}/\Esp{\Zn}$$

Again, no  general formula can be  computed and the  expression of the
speed-up will depend on the distribution of $Y$.

However, it  is worth noting that  our computation of  the speed-up is
related to 
order   statistics,   see   \cite{david2003order}   for   a   detailed
presentation.  
For  instance, the  first order  statistics of  a distribution  is its
minimal   value,   and   the   $k^{th}$   order   statistic   is   its
$k^{th}$-smallest value.   For predicting  the speedup, we  are indeed
interested  in computing the  expectation of  the distribution  of the
minimum  draw.   As the  above  formula  suggests,  this may  lead  to
heavy    calculations,     but    recent    studies     such    as
\cite{Nadarajah2008}  give  explicit formulas  for  this quantity  for
several classical probability distributions.

\subsection{Case of an Exponential Distribution}
\label{exponential-distribution}

Assume that $Y$ has a shifted exponential distribution, as it has been
suggested by~\cite{DBLP:journals/heuristics/AiexRR02,tttplots2007}.
\begin{eqnarray*}
\Distrib{Y}(t)=& \left\{
\begin{array}{l l}
0 & \text{ if } t< x_0\\
  \lambda e^{-\lambda (t-x_0)} &\text{ if } t>x_0 \\
  \end{array}
  \right.
  \\
\cumulative{Y}(t)=& \left\{
\begin{array}{l l}
0 & \text{ if } t< x_0\\
1-  e^{-\lambda (t-x_0)} &\text{ if } t>x_0 \\
  \end{array}
  \right.
  \\
  \Esp{Y}=&x_0+1/\lambda
\end{eqnarray*}
Then the above formula can be symbolically computed by hand:
\begin{eqnarray*}
\Distrib{\Zn}(t)=& \left\{
\begin{array}{l l}
0 & \text{ if } t< x_0\\
 n \lambda e^{-n \lambda (t-x_0)} &\text{ if } t>x_0 \\
  \end{array}
  \right.
  \\
\cumulative{\Zn}(t)=& \left\{
\begin{array}{l l}
0 & \text{ if } t< x_0\\
1-  e^{-n \lambda (t-x_0)} &\text{ if } t>x_0 \\
  \end{array}
  \right.
\end{eqnarray*}

\begin{figure}[t]
\begin{center}
\includegraphics[width=0.45\textwidth]{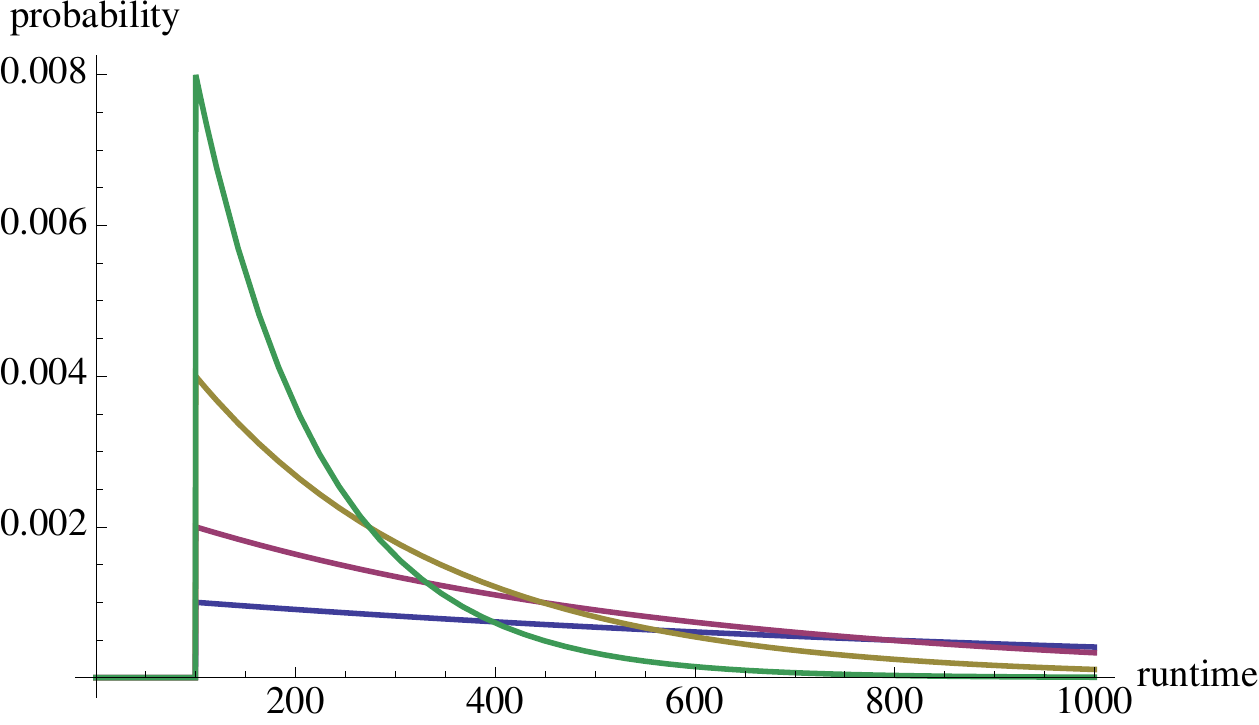}
\caption{\label{lois-exponentielle}For  an  exponential  distribution,
  here in blue with $x_0=100$ and $\lambda=1/1000$, simulations of the
  distribution  of \Zn  for  $n=2$ (pink),  $n=4$  (yellow) and  $n=8$
  (green).}
\end{center}
\end{figure}

The intuitive observation  of section \ref{Min-Distribution} is easily
seen  on the  expression of  the parallel  distribution, which  has an
initial value  multiplied by $n$ but an  exponential factor decreasing
$n$-times     faster,     as     shown     on    the     curves     of
Figure~\ref{lois-exponentielle}.

And in this case, one  can symbolically compute both the expectation
and speed-up for \Zn:
\begin{eqnarray*}
\Esp{\Zn}&=& n \lambda \int_{x_0}^\infty t e^{- n \lambda (t-x_0)} dt \\
&=& x_0+\frac{1}{n\lambda} \\
  \Gain{n}&=& \frac{x_0+\frac{1}{\lambda} }{x_0+\frac{1}{n\lambda}} \\
\end{eqnarray*}

\begin{figure}[t]
\begin{center}
\includegraphics[width=0.4\textwidth]{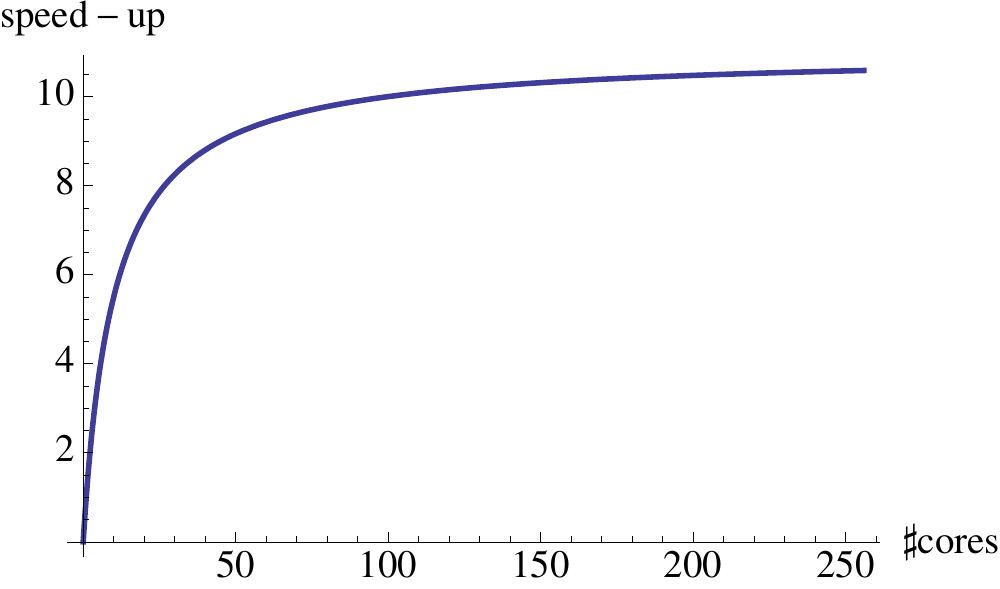}
\caption{\label{speedup-exponentielle}Predicted speed-up    in   case
of an exponential  distribution,   with  $x_0=100$  and  $\lambda=1/1000$,
w.r.t. the number of cores.}
\end{center}
\end{figure}

Figure~\ref{speedup-exponentielle} shows the evolution of the speed-up
when the number of cores  increases. With such a rather simple formula
for the speed-up, it is worth studying what happens when the number of
cores $n$ tends to infinity.  Depending on the chosen algorithm, $x_0$
may be null or not. If  $x_0=0$, then the expectation tends to $0$ and
the  speed-up is  equal to  $n$. This  case has  already  been studied
by~\cite{Aarts95}.  For  $x_0>0$, the  speed-up admits a  finite limit
which  is $\frac{x_0+\frac{1}{\lambda}}{x_0}=1+ \frac{1}{x_0\lambda}$.
Yet, this  limit may be  reached slowly, but  depends on the  value of
$x_0$ and $\lambda$:  obviously, the closest $x_0$ is  to zero and the
higher it  will be.  Another  interesting value is the  coefficient of
the tangent at the origin, which approximates the speed-up for a small
number   of    cores.   In   case    of   an   exponential,    it   is
$(x_0*\lambda+1)$. The  higher $x_0$ and $\lambda$, the  bigger is the
speed-up  at  the beginning.  In  the  following,  we will  see  that,
depending  on  the  combinations  of $x_0$  and  $\lambda$,  different
behaviors can be observed.


\subsection{Case of a Lognormal Distribution}
\label{lognormal-disitribution}

Other distributions  can be considered,  depending on the  behavior of
the base algorithm. We will study the case of a lognormal distribution,
which is  the log of a  gaussian distribution, because it will be shown
in Section~\ref{MS200} that it fits one experiment. It  has two parameters,
the mean $\mu$ and the standard deviation $\sigma$. In the same way as
the shifted exponential,  we shift the distribution so  that it starts
at  a given  parameter $x_0$.  Formally, a (shifted) lognormal  distribution is
defined as:

\begin{eqnarray*}
\Distrib{Y}(t)=& \left\{
\begin{array}{l l}
0 & \text{ if } t< x_0\\
\frac{e^{-\frac{(-\mu+log(t-x_0))^2}{2\sigma^2}}}{\sqrt{2\pi} (t-x_0) \sigma} &\text{ if } t>x_0 \\
  \end{array}
  \right.
  \\
\cumulative{Y}(t)=& \left\{
\begin{array}{l l}
0 & \text{ if } t< x_0\\
\frac{1}{2} \text{erfc}(\frac{\mu-log(t-x_0)}{\sqrt{2}\sigma}) &\text{ if } t>x_0 \\
  \end{array}
  \right.
\end{eqnarray*}

where   erfc  is   the   complementary  error   function  defined   by
$\text{erfc}(z)=\frac{2}{\sqrt{\pi}}\int_z^\infty e^{-t^2}
dt$. 

\begin{figure}[t]
\begin{center}
\includegraphics[width=0.4\textwidth]{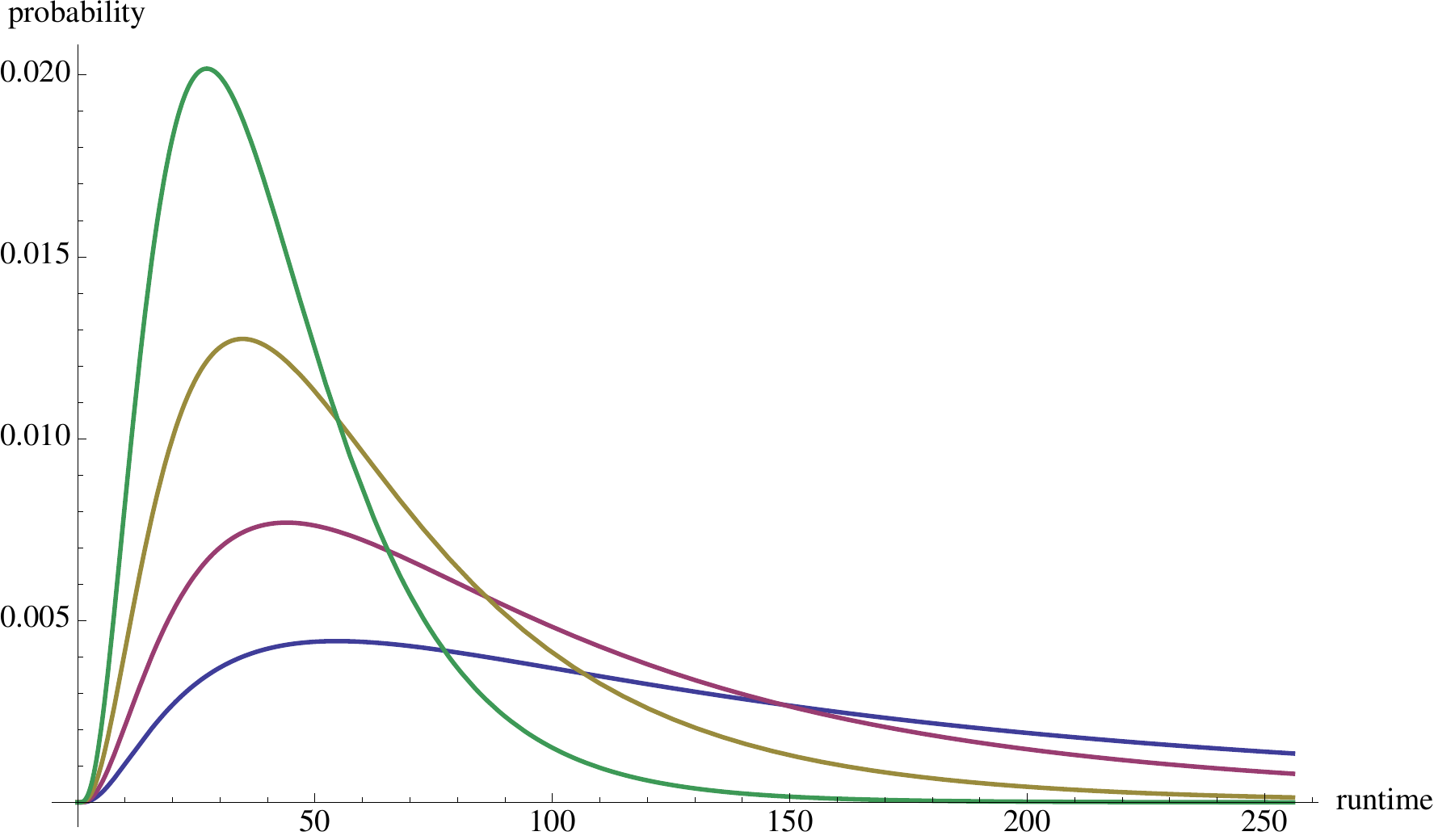}
\caption{\label{loi-lognormale}For  a lognormal distribution,  here in
  blue  with  $x_0=0$,  $\mu=5$  and $\sigma=1$,  simulations  of  the
  distribution  of \Zn  for  $n=2$ (pink),  $n=4$  (yellow) and  $n=8$
  (green).}
\end{center}
\end{figure}

\begin{figure}[t]
\begin{center}
\includegraphics[width=0.4\textwidth]{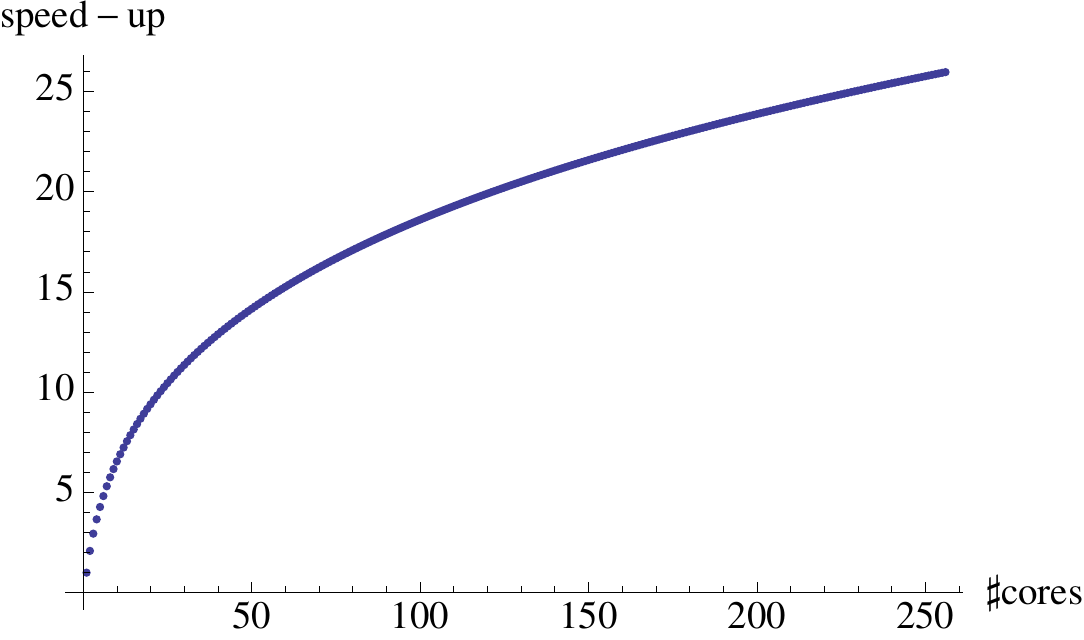}
\caption{\label{speedup-lognormale}Predicted speed-up  in  case  of a  lognormal
  distribution, with $x_0=0$, $\mu=5$ and $\sigma=1$, depending on the
  number of cores (on the abscissa).}
\end{center}
\end{figure}

Figure~\ref{loi-lognormale}  depicts lognormal  distributions  of \Zn,
for several  $n$. The  computations for the  distribution of  \Zn, its
expectation  and  the   theoretical  speed-up  are  quite  complicated
formulas.  But \cite{Nadarajah2008} gives  an explicit formula for all
the  moments  of lognormal  order  statistics  with  only a  numerical
integration  step, from  which  we  can derive  a  computation of  the
speed-up (since the  expectation of \Zn is the  first order moment for
the first order statistics).  This allows us to draw the general shape
of     the     speed-up,     an     example     being     given     on
Figure~\ref{speedup-lognormale}.   Due  to  the numerical  integration
step, which requires numerical values  for the number of cores $n$, we
restrict  the  computation  to  integer  values of  $n$.   This  is  a
reasonable limitation as the number of cores is indeed an integer.

\section{Application to Local Search}
\label{local-search}

Since  about a decade,  the interest  for the  family of  Local Search
methods  and Metaheuristics for  solving large  combinatorial problems
has  been growing  and  has  attracted much  attention  from both  the
Operations  Research and the  Artificial Intelligence  communities for
solving                                                       real-life
problems~\cite{handbook-approx,comet05,metaheuristics}.

\subsection{Local Search for Constraint Solving}

Local Search starts  from a random configuration and  tries to improve
this  configuration, little by  little, through  small changes  in the
values of the problem variables.  Hence  the term ``local search'' as, at each
time  step, only  new  configurations that  are  ``neighbors'' of  the
current   configuration  are   explored.   The   definition   of  what
constitutes a  neighborhood will  of course be  problem-dependent, but
basically it  consists in changing the  value of a  few variables only
(usually one or  two).  The advantage of Local  Search methods is that
they  will  usually  quickly  converge  towards  a  solution  (if  the
optimality  criterion  and  the  notion of  neighborhood  are  defined
correctly...) and not exhaustively explore the entire search space.

Applying Local  Search to  Constraint Satisfaction Problems  (CSP) has
been    attracting    some    interest    since   about    a    decade
\cite{DBLP:conf/saga/CodognetD01,Hao00,comet05}, as it can tackle CSPs
instances far beyond the  reach of classical propagation-based
constraint solvers \cite{PVH89,Bessiere06}.
A generic, domain-independent  constraint-based  local  search   method,
named Adaptive  Search, has been  proposed by 
\cite{DBLP:conf/saga/CodognetD01,mic/CodognetD03}.   
This meta-heuristic  takes advantage  of the  structure of  the  problem in
terms  of constraints  and variables  and  can guide  the search  more
precisely than a single global  cost function to optimize, such as for
instance the number of  violated constraints.  The algorithm also uses
a short-term adaptive memory in the  spirit of Tabu Search in order to
prevent stagnation in local minima and loops.

\subsection{Adaptive Search}

An implementation of Adaptive Search (AS) has been developed in C language 
as a framework library and is available as a freeware at the URL:\\
\mbox{\url{http://cri-dist.univ-paris1.fr/diaz/adaptive/}}

We used this reference implementation for our experiments.
The  Adaptive  Search  method can  be  applied  to  a large  class  of
constraints   (\textit{e.g.}    linear   and   non-linear   arithmetic
constraints,  symbolic constraints,  etc.)  and  naturally  copes with
over-constrained problems  \cite{Truchet04}.  The input  of the method
is a Constraint Satisfaction Problem (CSP for short), which is defined
as a  triple (X;D;C), where  X is a  set of variables,  D is a  set of
domains, i.e.,  finite sets  of possible values  (one domain  for each
variable), and C a set  of constraints restricting the values that the
variables can simultaneously take. For each constraint, an \emph{error
  function} needs to be defined;  it gives, for each tuple of variable
values, an  indication of how  much the constraint is  violated.  This
idea has also been proposed independently by \cite{Hao00}, where it is
called  ``penalty functions'',  and then  reused by  the  Comet system
\cite{comet05}, where  it is called ``violations''.   For example, the
error function associated with an arithmetic constraint $|X - Y| < c$,
for a given constant $c \geq 0$, can be $max(0, |X-Y|-c)$.

Adaptive
Search relies  on iterative repair,  based on variable  and constraint
error information, seeking  to reduce the error on  the worst variable
so far.   The basic  idea is  to compute the  error function  for each
constraint,  then  combine  for   each  variable  the  errors  of  all
constraints in which it  appears, thereby projecting constraint errors
onto  the   relevant  variables.    This  combination  of   errors  is
problem-dependent,  see \cite{DBLP:conf/saga/CodognetD01}  for details
and examples,  but it  is usually a  simple sum  or a sum  of absolute
values, although  it might also be  a weighted sum  if constraints are
given different  priorities.  Finally,  the variable with  the highest
error is designated as the  ``culprit'' and its value is modified.  In
this   second   step,    the   well   known   min-conflict   heuristic
\cite{min-conflict} is used to select the value in the variable domain
which is  the most promising, that  is, the value for  which the total
error in the next configuration is minimal.

In order to prevent being
trapped in  local minima, the  Adaptive Search method also  includes a
short-term   memory  mechanism  to   store  configurations   to  avoid
(variables  can  be  marked  Tabu  and  ``frozen''  for  a  number  of
iterations).    It  also  integrates   reset  transitions   to  escape
stagnation around  local minima.  A reset consists  in assigning fresh
random values  to some variables  (also randomly chosen).  A  reset is
guided  by the  number of  variables being  marked Tabu.   It  is also
possible to restart from scratch when the number of iterations becomes
too large (this  can be viewed as  a reset of all variables  but it is
guided  by the  number of  iterations).   The core  ideas of  adaptive
search can be summarized as follow:
\begin{itemize}
\item to  consider for  each constraint a  heuristic function  that is
  able to compute an approximated  degree of satisfaction of the goals
  (the current \emph{error} on the constraint);
\item to aggregate constraints on  each variable and project the error
  on variables  thus trying to  repair the \emph{worst}  variable with
  the most promising value;
\item  to keep  a short-term  memory  of bad  configurations to  avoid
  looping (\textit{i.e.}  some sort of \emph{tabu list}) together with
  a reset mechanism.
\end{itemize}

\section{Benchmark Problems and Experimental Results}
\label{benchmarks}

We have  chosen to  test this  method on two  problems from  the CSPLib
benchmark library~\cite{CSPLIB},  and on a  hard combinatorial problem
abstracted   from  radar  and   sonar  applications.    After  briefly
introducing the  classical benchmarks,  we detail the  latter problem,
called \ca.  Then  we show the performance and  the speed-ups obtained
with  both sequential and  a multi-walk  Adaptive Search  algorithm on
these problems.

We  use two classical  benchmarks from  CSPLib
consisting  of:
\begin{itemize}
\item The \ai Series problem (prob007 in CSPLib),
\item The \ms problem (prob019 in CSPLib).
\end{itemize}

Although  these  benchmarks are  academic,  they  are abstractions  of
real-world problems and could  involve very large combinatorial search
spaces, \eg  the 200$\times$200 \ms problem  requires 40,000 variables
whose domains range over 40,000 values. Indeed the search space in the
Adaptive Search model (using permutations) is $40,000!$, \ie more than
$10^{166713}$ configurations.
Classical  propagation-based  constraint  solvers  cannot  solve  this
problem  for instances  higher  than  20x20.  Also  note  that we  are
tackling  constraint  {\it   satisfaction}  problems  as  optimization
problems, that is, we want  to minimize the global error (representing
the  violation of  constraints)  to value  zero,  therefore finding  a
solution  means  that  we  actually  reach the  bound  (zero)  of  the
objective function to minimize.


\subsection{The \ai Series Problem}
\label{sec:all-interval-series}

\includegraphics[width=0.48\textwidth]{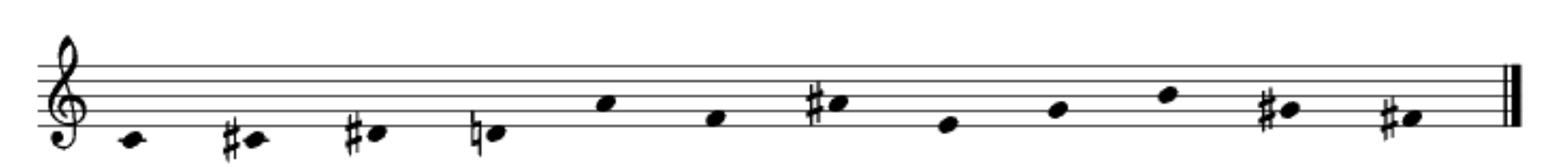}

\noindent This problem is described as \texttt{prob007} in the CSPLib.
This    benchmark   is    in   fact    a   well-known    exercise   in
music~\cite{DBLP:journals/soco/TruchetC04}   where  the  goal   is  to
compose a sequence of $N$ notes  such that all are different and tonal
intervals between consecutive notes are also distinct. This problem is
equivalent to  finding a  permutation of the  $N$ first  integers such
that the absolute difference  between two consecutive pairs of numbers
are all different.  This amounts to finding a permutation $(X1, \ldots
X_N)$  of  $\{0, \ldots  N-1\}$  such  that  the list  $(abs(X_1-X_2),
abs(X_2-X_3)   \ldots  abs(X_{N-1}-   x_N))$  is   a   permutation  of
${1,\ldots,N-1}$.    A   possible   solution    for   $N   =   8$   is
$(3,6,0,7,2,4,5,1)$ since all consecutive distances are different:

\newcommand{\Dist}[1]{$\frac{#1}{~~~~}$}
\begin{center}
3 \Dist{3} 6 \Dist{6} 0 \Dist{7} 7 \Dist{5} 2 \Dist{2} 4 \Dist{1} 5 \Dist{4} 1
\end{center}


\subsection{The \ms Problem}
\label{sec:magic-squares}

\parpic[l]{
\begin{tabular}{rl|@{}c@{}|@{}c@{}|@{}c@{}|@{}c@{}|@{}l}
\cline{3-6}
   &            &   16        &     3        &     2        &    13        & $~\rightarrow$ 34 \\ \cline{3-6}
   &            &    5        &    10        &    11        &     8        & $~\rightarrow$ 34 \\ \cline{3-6}
   &            &    9        &     6        &     7        &    12        & $~\rightarrow$ 34 \\ \cline{3-6}
   &            &    4        &    15        &    14        &     1        & $~\rightarrow$ 34 \\ \cline{3-6}
   &\NN{$\swarrow$}&\N{$\downarrow$}&\N{$\downarrow$}&\N{$\downarrow$}&\N{$\downarrow$}&\multicolumn{1}{@{}l}{$~\searrow$}       \\
\multicolumn{2}{@{}l}{34}    &\N{34}       & \N{34}       & \N{34}       & \N{34}       & \multicolumn{1}{r}{34}
\end{tabular}
}

\noindent The \ms problem  is catalogued as \texttt{prob019} in CSPLib
and  consists  in placing  the  numbers  $\{1,2  \cdots N^2\}$  on  an
$N\times N$ square such that  each row, column and main diagonal equal
the same sum (the constant $N(N^2+1)/2$).

For  instance, this figure  shows a  well-known solution  for $N  = 4$
(depicted by  Albrecht D\"{u}rer in  his engraving \textit{Melancholia
  I}, 1514).

\subsection{The \ca Problem}
\label{sec:cap}

\parpic[l]{
  \includegraphics[width=3cm]{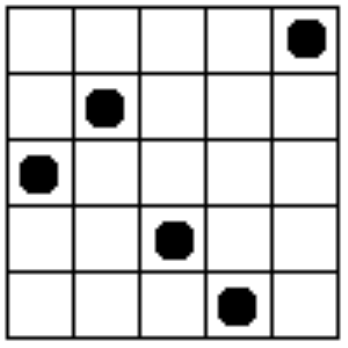}
}

\noindent A Costas array is an  $N \times N$ grid containing $N$ marks
such that  there is exactly  one mark per  row and per column  and the
$N(N -  1)/2$ vectors  joining the marks  are all different.   We give
here an  example of Costas array of  size 5.  It is  convenient to see
the \ca Problem (CAP) as a permutation problem by considering an array
of $N$ variables $(V_1,\ldots,V_N)$ which forms a permutation of $\{1,
2, \ldots, N\}$. The above  Costas array can thus be represented by
the array $[3,4,2,1,5]$.

Historically these arrays have been developed in the 1960's to compute
a set of  sonar and radar frequencies avoiding  noise \cite{Cos-84}. A
very complete survey  on Costas arrays can be  found in \cite{Dra-06}.
The problem of finding a Costas array of size $N$  is very complex since
the  required  time grows  exponentially  with  $N$.   In the  1980's,
several algorithms  have been proposed  to build a Costas  array given
$N$ (methods to  produce Costas arrays of order 24 to  29 can be found
in~\cite{4285351,n28,n29,DBLP:conf/ciss/RussoEB10}), such as the Welch
construction \cite{Gol-84}  and the Golomb  construction \cite{GT-84},
but these  methods cannot  built Costas arrays  of size $32$  and some
higher non-prime sizes.  Nowadays,  after many decades of research, it
remains unknown if there exist any Costas arrays of size $32$ or $33$.
Another  difficult problem  is to  enumerate all  Costas arrays  for a
given  size.   Using  the  Golomb and  Welch  constructions,  Drakakis
\emph{et.   al} present in  \cite{n29} all  Costas arrays  for $N=29$.
They show that among the $29!$ permutations, there are only 164 Costas
arrays, and 23 unique Costas arrays up to rotation and reflection.

\subsection{Sequential Results}
\label{sec:seq-results}

We run  our benchmarks in a  sequential manner in order  to have about
650 runtimes  for each.  Sequential  experiments, as well  as parallel
experiments,  have been  done  on the  \emph{Griffon}  cluster of  the
Grid'5000    platform.     The   following    Tables~\ref{tab:seqtime}
and~\ref{tab:seqiter}  shows  the minimum,  mean,  median and  maximum
respectively among  the runtimes and  the number of iterations  of our
benchmarks.

\begin{table}[!h]
  \small
  \centering
  \begin{tabular}{||c|r|r|r|r||}%
    \hline%
    \CF{Problem} & \C{Min}  & \C{Mean}  & \C{Median}  & \D{Max} \\
    \hline\hline
    MS~200 & 5.51 & 382.0 & 126.3 & 7,441.6\\
    AI~700 & 23.25 & 1,354.0 & 945.4 & 10,243.4\\
    Costas~21 & 6.55 & 3,744.4 & 2,457.4 & 19,972.0\\
    \hline%
  \end{tabular}
  \caption{Sequential execution times (in seconds)}
  \label{tab:seqtime}
\end{table}

\begin{table}[!h]
  \small
  \centering
  \begin{tabular}{||c|r|r|r|r||}%
    \hline%
    \CF{Problem} & \C{Min}  & \C{Mean}  & \C{Median}  & \D{Max} \\
    \hline\hline
    MS~200 & 6,210 & 443,969 & 164,042 & 7,895,872\\
    AI~700 & 1,217 & 110,393 & 76,242 & 826,871\\
    Costas~21 & 321,361 & 183,428,617 & 119,667,588 & 977,709,115\\
    \hline%
  \end{tabular}
  \caption{Sequential number of iterations}
  \label{tab:seqiter}
\end{table}

One can see  that runtimes and the number  of iterations, respectively
from Tables~\ref{tab:seqtime} and~\ref{tab:seqiter}, are spread over a
large  interval,  illustrating  the  stochasticity of  the  algorithm.
Depending  on the  benchmark,  there is  a  ratio of  a few  thousands
between the minimum and the maximum runtimes.

\subsection{Parallel Results}
\label{sec:par-results}


We    have   conduct    parallel   experiments    on    the   Grid5000
platform~\cite{Grid5000}, the French national grid for research, which
contains 8,596 cores  deployed on 11 sites distributed  in France. For
our experiments, we used the \emph{Griffon} cluster at Nancy, composed
of  184 Intel  Xeon L5420  (Quad-core, 2.5GHz,  12MB of  L2-cache, bus
frequency at 1333MHz), thus with a total of 736 cores available giving
a peak performances of 7.36TFlops.


Tables~\ref{tab:benchstime}   and~\ref{tab:benchsiter}   present   the
execution  times  and the  number  of  iterations, respectively,  with
speed-ups for  executions of  large benchmarks \ms  200$\times$200, \ai
with $n$ =  700 and \ca with  $n=21$, up to 256 cores.   The same code
has been ported and executed, timings are given in seconds and are the
average  of  50  runs.   One  can  notice  there  are  no  significant
differences between  speed-ups of these two tables,  therefore we will
prefer as a time measurement the  number of iterations, which has the good
property of not being machine-dependent. Anyway, similar speed-ups have
been achieved on other parallel machines~\cite{ppopp12}.

\begin{table}[!h]
  \small
  \centering
  \begin{tabular}{||c|r|r|r|r|r|r||}%
    \hline%
    \CF{Problem} &time on 1 core& \multicolumn{5}{c||}{speed-up on $k$ cores} \\
    \cline{3-7}
    \CF{} & \C{(seconds)}  & \C{16}  & \C{32}  & \C{64} & \C{128} & \D{256} \\
    \hline\hline
    MS~200    & 382.0   & 18.3 &  24.5  & 32.3 & 37.0 & 47.8 \\
    AI~700 & 1,354.0 & 12.9 & 19.3 & 30.6 & 39.2 & 45.5 \\
    Costas~21 & 3,744.4 & 15.7 & 26.4 & 59.8 & 154.5 & 274.8 \\
    \hline%
  \end{tabular}
  \caption{Speed-ups with respect to sequential time}
  \label{tab:benchstime}
\end{table}

\begin{table}[!h]
  \small
  \centering
  \begin{tabular}{||c|r|r|r|r|r|r||}%
    \hline%
    \CF{Problem} & \# of iterations& \multicolumn{5}{c||}{speed-up on $k$ cores} \\
    \cline{3-7}
    \CF{} & \C{on 1 core}  & \C{16}  & \C{32}  & \C{64} & \C{128} & \D{256} \\
    \hline\hline
    MS~200    & 443,969   & 16.6 &  22.2  & 29.9 & 34.3 & 45.0 \\
    AI~700 & 110,393 & 12.8 & 20.2 & 29.3 & 37.3 & 48.0 \\
    Costas~21 & 183,428,617 & 15.8 & 26.4 & 60.0 & 159.2 & 290.5 \\
    \hline%
  \end{tabular}
  \caption{Speed-ups with respect to sequential number of iterations}
  \label{tab:benchsiter}
\end{table}

\begin{figure}[h!]
  \centering
  \includegraphics[width=\linewidth]{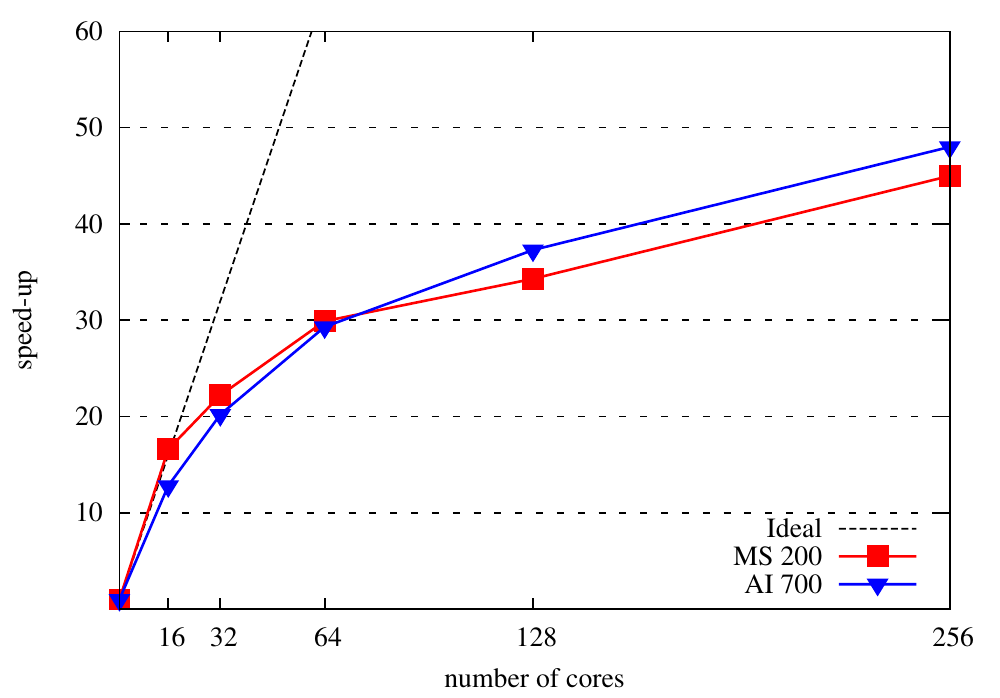}
  \caption{Speed-ups for CSPLib benchmarks}
  \label{fig:benchs}
\end{figure}
\begin{figure}[h!]
  \centering
  \includegraphics[width=\linewidth]{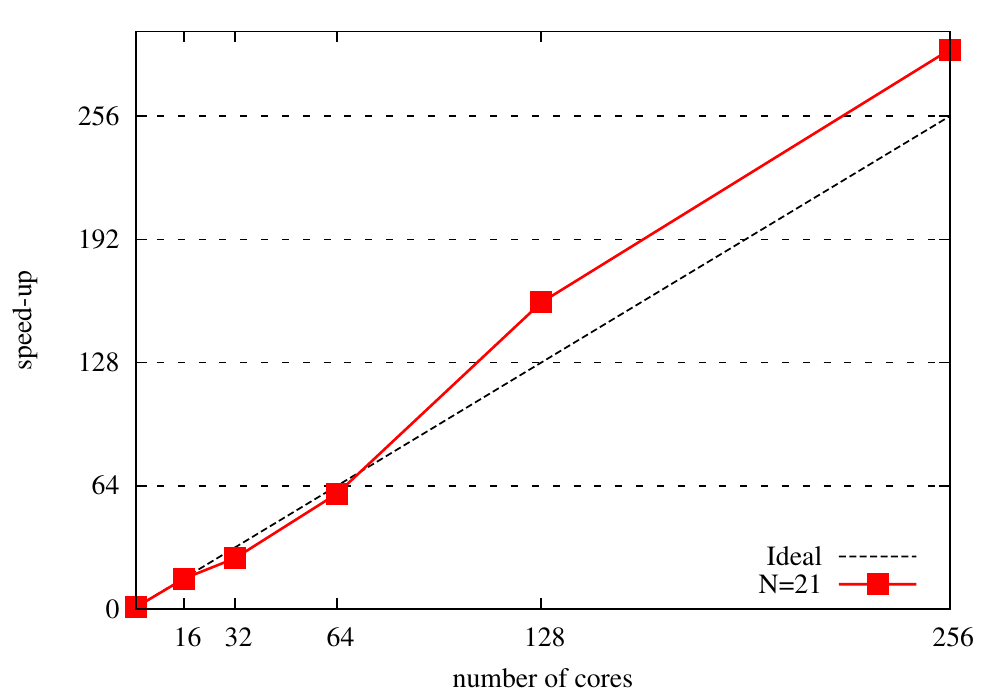}
  \caption{Speed-ups for the \ca Problem}
  \label{fig:costas}
\end{figure}

For the two CSPLib benchmarks, one can observe the stabilization point
is not yet  obtained for 256 cores, even if  speed-ups do not increase
as fast  as the  number of  cores, \ie are  getting further  away from
linear     speed-up.      This     is     visually     depicted     on
Figure~\ref{fig:benchs}.  For the \ca  Problem, the AS algorithm reaches
linear or  even supra-linear speed-ups  up to 256 cores, as depicted in
Figure~\ref{fig:costas}.
Actually, it scales linearly  far beyond this point, \ie at  least up to
8,192    cores, as reported in~\cite{pco12}.


\section{Prediction of Parallel Speed-ups}
\label{prediction}

On each  problem, the sequential  benchmark gives observations  of the
distribution of  the algorithm  runtime $\Distrib{Y}$. Yet,  the exact
distribution is  still unknown. It  can be any real  distribution, not
even a classical one. In the following, we will rely on the assumption
that  $Y$ is  distributed with  a known  parametric  distribution.  We
perform  a statistical  test, called  Kolmogorov-Smirnov test,  on the
hypothesis $\mathcal{H}_0$ that  the collected observations correspond
to  a  theoretical distribution.  Assuming  $\mathcal{H}_0$, the  test
first computes the probability that the distance between the collected
data and  the theoretical distribution, does  not significantly differ
from its theoretical value. This probability is called the p-value.
Then, the p-value is compared  to a fixed threshhold (usually $0.05$).
If it is  smaller, one rejects $\mathcal{H}_0$. For  us, it means that
the observations  do not  correspond to the  theoretical distribution.
If the p-value is high, we  will consider that the distribution of $Y$
is  approximated  by  the  theoretical  one. Note that the
Kolmogorov-Smirnov test is  a statistical test, which in  no way proves
that $Y$ follows the distribution.   However, it measures how well the
observations fit  a theoretical curve and,  as it will be  seen in the
following, it is accurate enough for our purpose.

The  distributions   tested  for  local  search   algorithms  are  the
exponential distribution,  as suggested by \cite{eadie1971statistical}
, and  the lognormal distribution, because  it appears to  fit the \ms
problem.  We have also  performed the Kolmogorov-Smirnov test on other
distributions  (e.g.  gaussian and  L{\'e}vy),  but obtained  negative
results  w.r.t. the experimental  benchmarks, thus  we do  not include
them in  the sequel.  For each  problem, we will need  to estimate the
value of the  parameters of the distribution, which is  done on a case
by case basis.

Once we  have an  estimated distribution for  the runtimes of  $Y$, it
becomes possible to compute the expectation of the parallel runtimes
and      the       speed-up      thanks      to       formulas      of
Section~\ref{expectation-and-speedup}.      All    the    mathematical
calculations  are done  with  Mathematica, a  commercial software  for
symbolic computation~\cite{mathematica}.

In  the  following,  all  the  analyses    are done  on  the  number  of
iterations,  because they  are more  likely  to be  unbiased.

\subsection{The \ai Series Problem}

The analysis is done on $720$ runs of the Adaptive Search algorithm on
the  instance  of  \ai  series  for  $700$  notes.   The  sequence  of
observations is written AI~700 in the following.

We  test  the  hypothesis   that  the  observations  admit  a  shifted
exponential        distribution         as        introduced        in
Section~\ref{exponential-distribution}.  The  first  step consists  in
estimating  the parameter  of the  distribution, which  for  a shifted
exponential  are  the  value   of  the  shift  $x_0$ and
$\lambda$\footnote{All  the notations are the same as in section
\ref{probabilistic-model}.}.
We  take for $x_0$ the minimum  observed value, $x_0=1217$.
The  exponential  parameter  is  estimated  thanks  to  the  following
relation: for a  non-shifted exponential distribution, the expectation
is            $1/\lambda$.           Thus            we           take
$\lambda=1/(\text{mean}(\text{AI~700})-x_0)$,        which        gives
$\lambda=9.15956*10^{-6}$.

\begin{figure}[t]
\begin{center}
\includegraphics[width=0.45\textwidth]{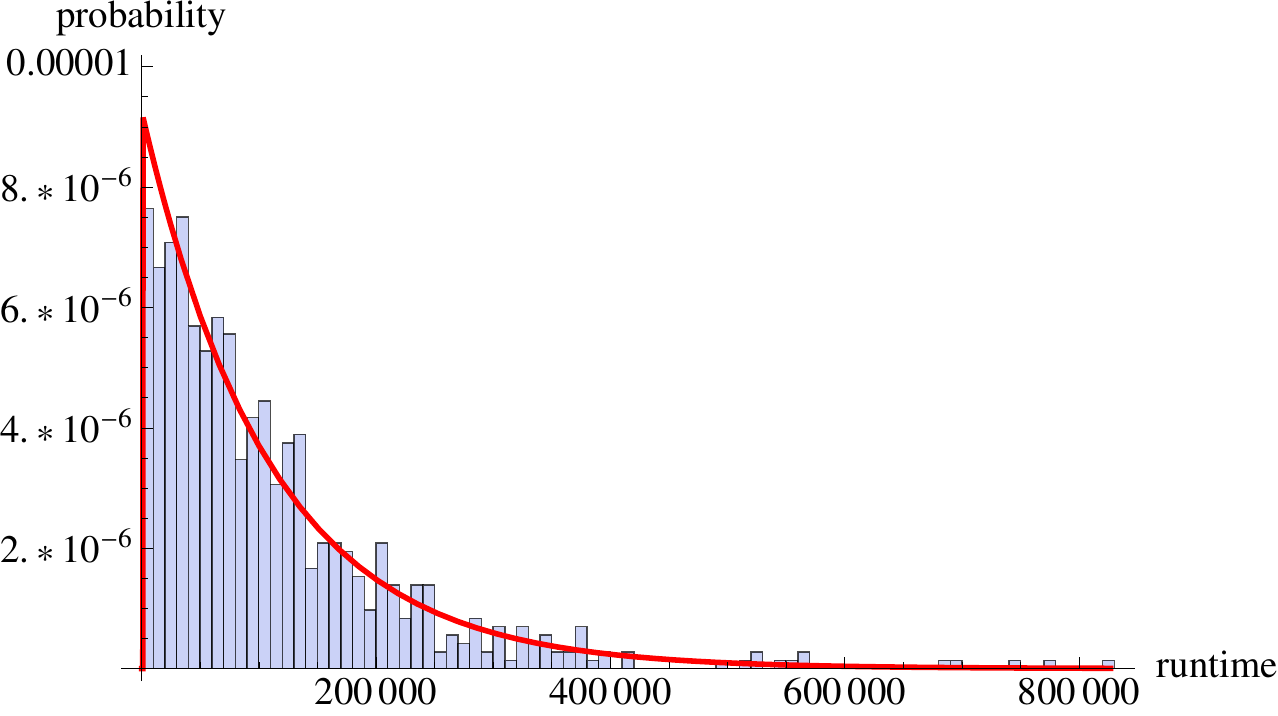}
\caption{\label{KS-AI700-exp}  Histogram  of  the observed  number  of
  iterations for $720$ runs on the \ai series problem with $N=700$, in
  blue.  In  red, the corresponding  shifted exponential distribution,
  statistically estimated.}
\end{center}
\end{figure}

We  then run the  Kolmogorov-Smirnov test  on the  shifted exponential
distribution with  these values of $x_0$ and  $\lambda$, which answers
positively   (computed  p-value:  $0.77435$).    We  thus   admit  the
hypothesis that AI~700 fits this shifted exponential distribution. As
an   illustration,  Figure~\ref{KS-AI700-exp}  shows   the  normalized
histogram of the observed runtimes and the theoretical distribution.

It is then  possible to symbolically compute the  speed-up that can be
expected     with     the     parallel     scheme     described     in
Section~\ref{multiwalk}.       We     use     the      formulas     of
Section~\ref{exponential-distribution}  with the  estimated parameters
and obtain a theoretical expression  for the speed-up.  This allows us
to calculate its value for different number of cores.

\begin{figure}[t]
  \begin{center}
  \includegraphics[width=0.45\textwidth]{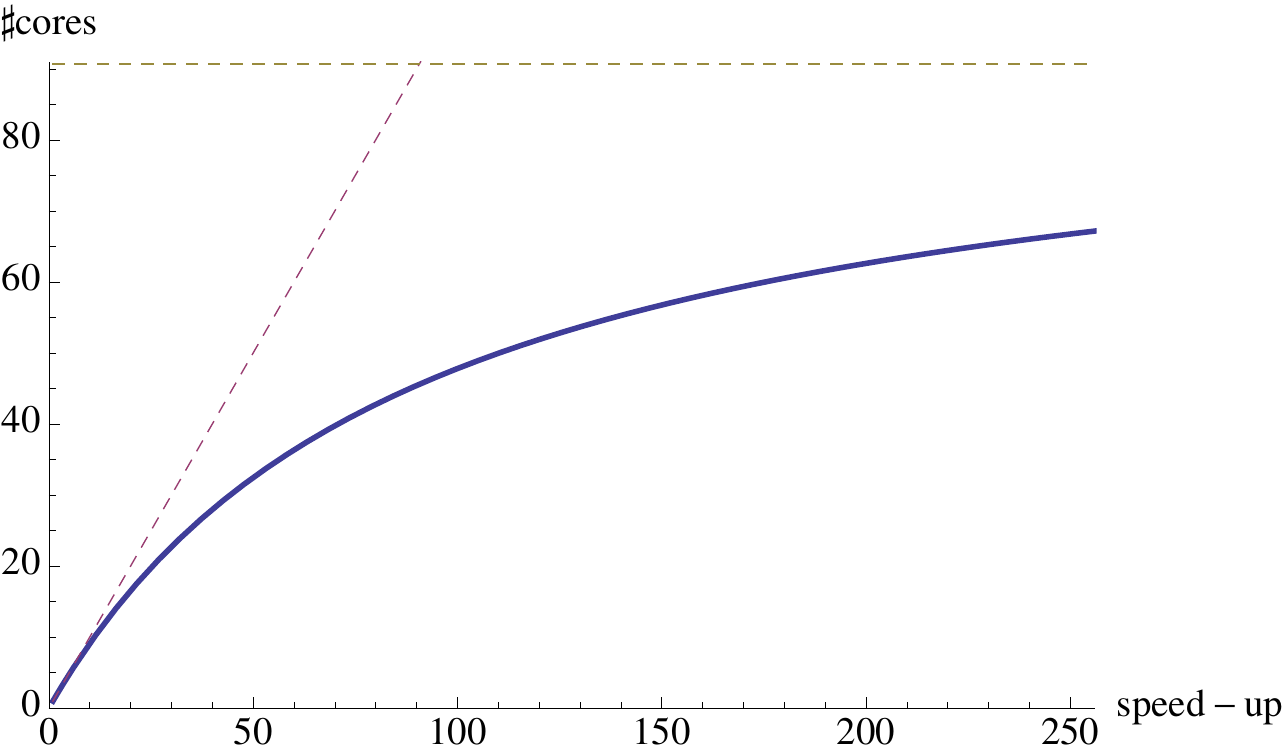}
  \caption{\label{speed-up-AI700} Predicted speed-up for  AI~700 as a function of
    the number of  cores (plain blue), with its  limit (dashed yellow)
    and the ideal linear speed-up (dashed pink).}
  \end{center}
\end{figure}

The  results  are  given  on Figure~\ref{speed-up-AI700}.   With  this
approximated distribution, the limit of  the speed-up when the number of
cores tends  to infinity is  $90.7087$. One can  see that, with  a 256
cores, the curve has not reached its limit, but comes close. Thus, the
speed-up  for this  instance of  \ai appears  significantly  less than
linear (linear meaning: equal to the number of cores).

\subsection{The \ms Series Problem}

\label{MS200}

\begin{figure}[t]
\begin{center}
\includegraphics[width=0.45\textwidth]{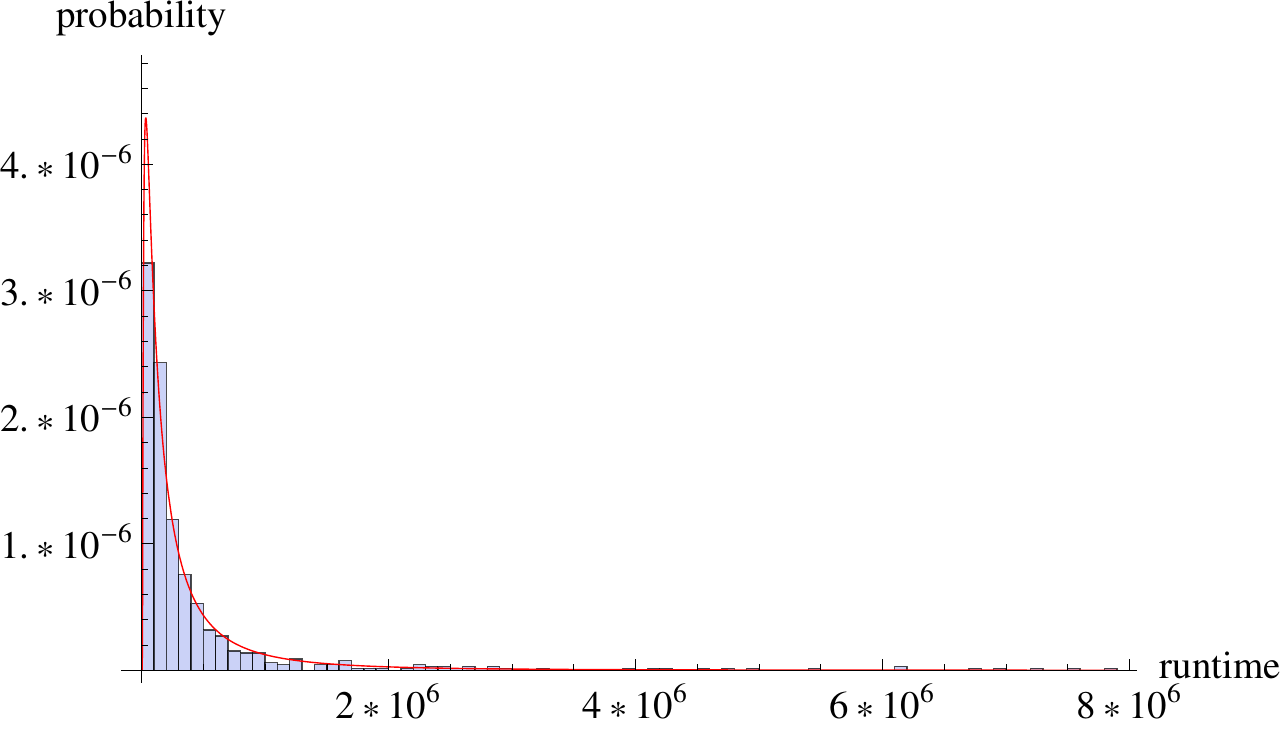}
\caption{\label{KS-MS200-lognormale} Histogram  of the observed number
  of iterations  for $662$  runs on the \ms problem with $N=200$, in
  blue.   In red,  the corresponding  shifted  lognormal distribution,
  statistically estimated.}
\end{center}
\end{figure}

For the \ms  problem with $N=200$, the observations  are the number of
iterations  on  $662$  runs,   with  a  minimum  of  $x_0=6210$.   The
Kolmogorov-Smirnov test  on a shifted  exponential distribution fails,
but we  obtain a positive  result with a lognormal  distribution, with
$\mu=12.0275$  and  $\sigma=1.3398$, shifted to $x_0$.
These parameters have been estimated with the use of the Mathematica software.
As an illustration, Figure~\ref{KS-MS200-lognormale}  shows the observations
and the theoretical estimated distribution.

\begin{figure}[t]
  \begin{center}
    \includegraphics[width=0.45\textwidth]{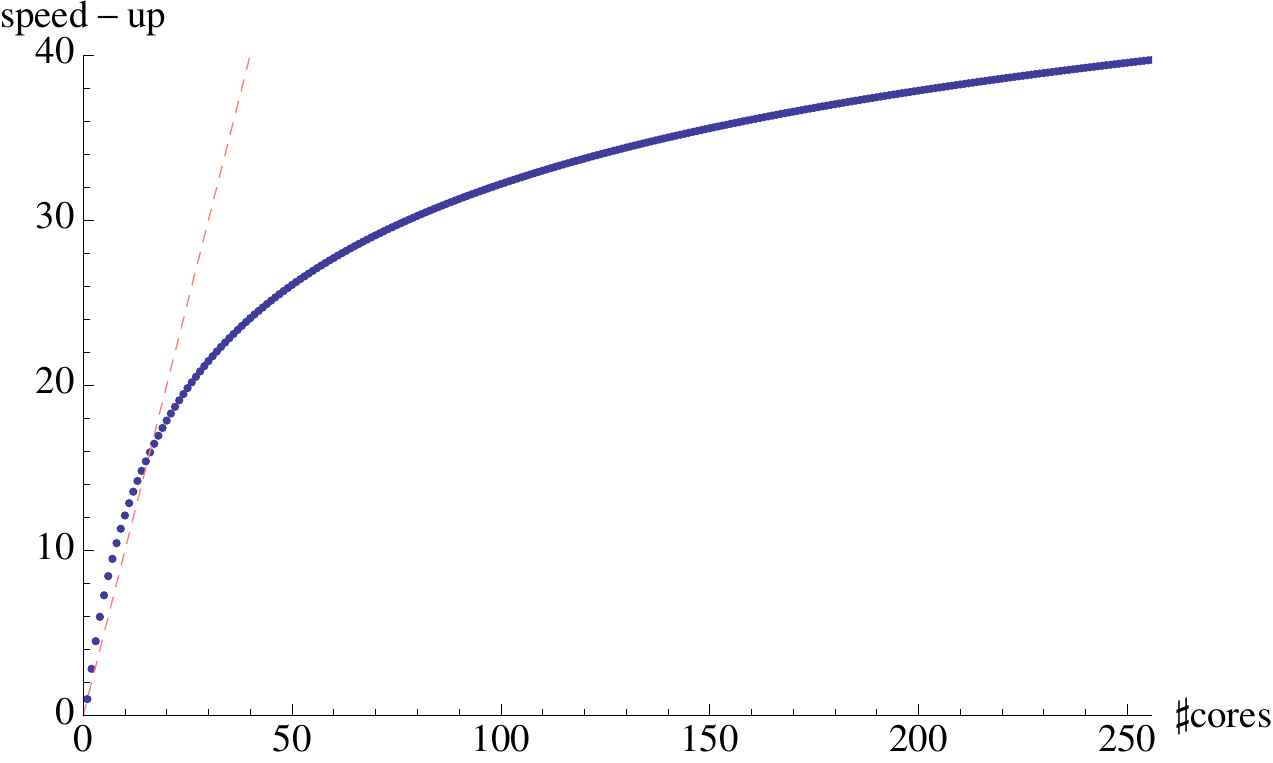}
    \caption{\label{speed-up-MS200} Predicted speed-up  for MS~200 as  a function
      of  the numbers  of cores  (blue  dots), with  the ideal  linear
      speed-up (dashed pink).}
  \end{center}
\end{figure}

The speed-up  can be computed by integrating  the minimum distribution
with numerical  integration techniques.  The results  are presented on
Figure~\ref{speed-up-MS200}.  We can observe  that the  speed-up grows
very fast  at the origin, which can  be explained by the  high peak of
the lognormal distribution with  these parameters. Again, the speed-up
is computed  with a numerical integration  step, and we  only draw the
curve  for  integer  values   of  $n$.
In this case again, the speed-up is significantly less than linear from
50 cores onwards, and the limit of the speed-up when the number of cores
tends to infinity is about 71.5.

\subsection{The \ca Problem}

The same analysis is done for the  runs of the AS algorithm on the \ca
problem  with $N=21$. The  observations are  taken from  the benchmark
with $638$ runs. The sequence of observations is written Costas~21.

\begin{figure}[t]
\begin{center}
\includegraphics[width=0.45\textwidth]{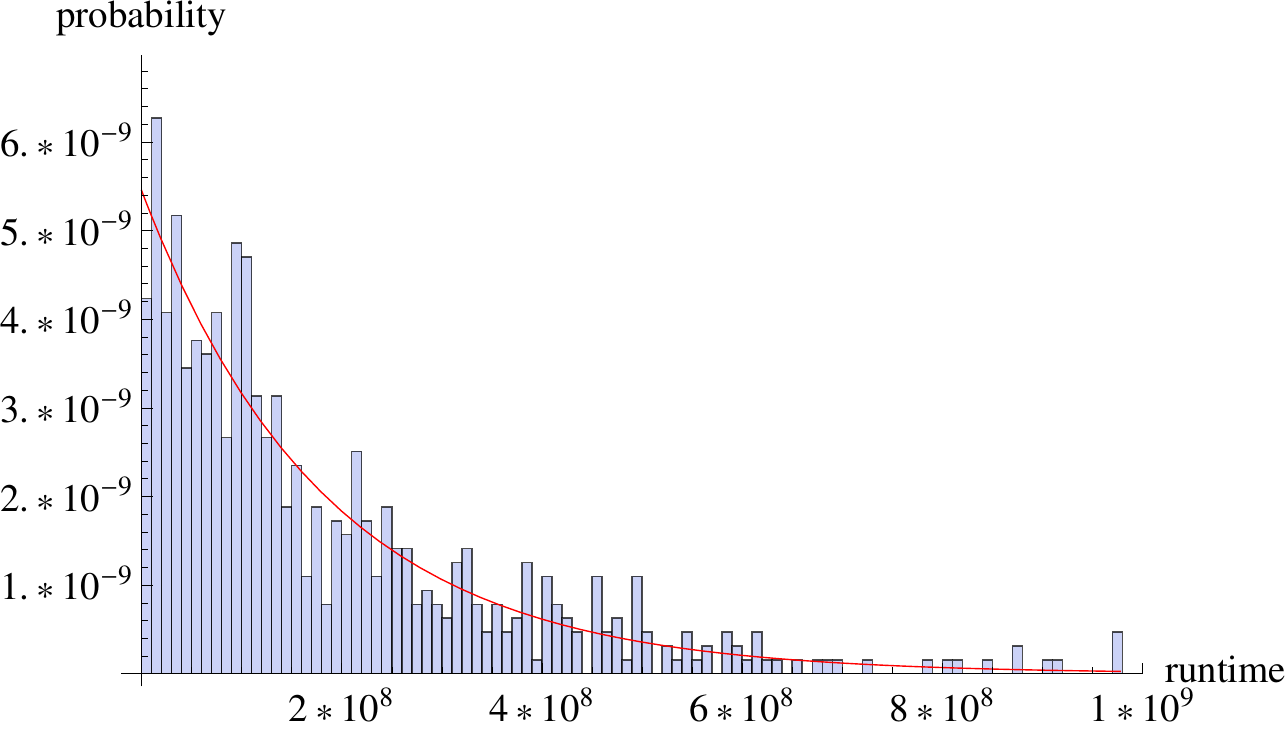}
\caption{\label{KS-COSTAS21-exp} Histogram  of the observed  number of
  iterations for $638$  runs on the \ca problem  with $N=21$, in blue.
  In red,  the corresponding shifted exponential distribution,
  statistically estimated.}
\end{center}
\end{figure}

This  benchmark has  an  interesting property:  the observed  minimum,
$3.2*10^5$ is neglictible compared  to its mean ($1.8*10^8$). Thus, we
estimate  $x_0=0$   and  perform  a  Kolmogorov-Smirnov   test  for  a
(non-shifted)          exponential          distribution,         with
$\lambda=1/\text{mean}(Costas~21)=5.4*10^{-9}$.   The test  is positive
for  this  exponential distribution,  with  a  p-value of  $0.751915$.
Figure~\ref{KS-COSTAS21-exp} shows the estimated distribution compared
to the observations.

\begin{figure}[t]
  \begin{center}
  \includegraphics[width=0.45\textwidth]{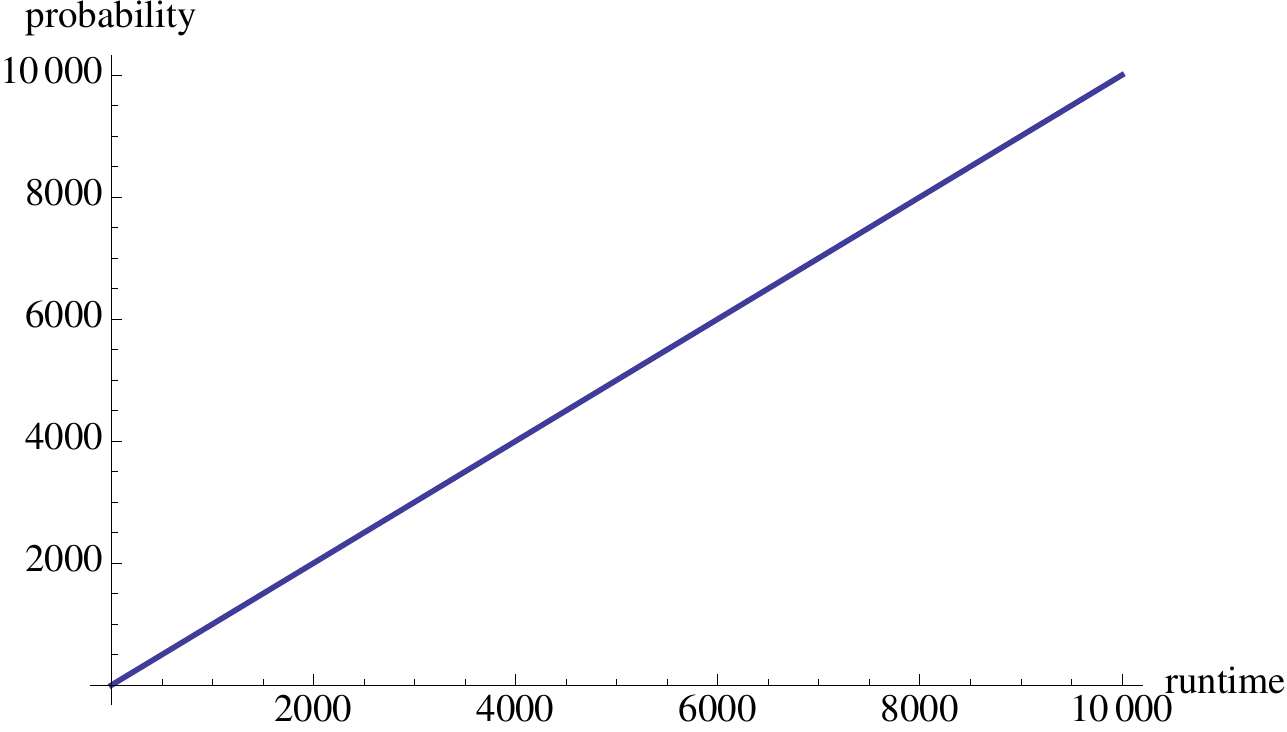}
  \caption{\label{speed-up-Costas21}   Predicted speed-up  for  Costas~21   as  a
    function of the number of cores.}
 \end{center}
\end{figure}

The computation of  the theoretical speed-up is also  done in the same
way as for  AI~700.  Yet, in this case, the  observed minimum for $x_0$
is  so  small  that  we   can  approximate  the  observations  with  a
non-shifted  distribution,  thus the  predicted  speed-up is  strictly
linear,  as   shown  in  Section~\ref{exponential-distribution}.   The
results  are given  on Figure~\ref{speed-up-Costas21}.   This explains
that one may observe linear speed-up when parallelizing \ca.

\section{Analysis}
\label{analysis}


Table~\ref{tab:comparison}   presents  the   comparison   between  the
predicted and the experimental speedups.  We can see that the accuracy
of  the prediction  is  very good up  to  64 parallel  cores  and then  the
divergence is limited even for 256 parallel cores.

For  the MS~200  problem,  the experimental  speed-up and the  predicted one
are almost identical up to 128 cores and diverging by 10\% for 256 cores.
For the AI~700 problem, the experimental speed-up is less good than the
predicted one by a maximum of 30\% for 128 and 256  cores.  For the Costas~21
problem, the experimental speed-up is better than the predicted one by
15\%  for  128 and  256  cores.

\begin{table}[!h]
  \small
  \centering
  \begin{tabular}{||c|r|r|r|r|r|r||}%
    \hline%
    \CF{Problem} & & \multicolumn{5}{c||}{speed-up on $k$ cores} \\
    \cline{3-7}
    \CF{} & \C{  }  & \C{16}  & \C{32}  & \C{64} & \C{128} & \D{256} \\
    \hline\hline
    MS~200     & experimental & 16.6 &  22.2  & 29.9 & 34.3 & 45.0 \\
              & predicted    & 15.94 & 22.04  & 28.28 & 34.26 & 39.7 \\
              &    &  &   & &  &  \\
    AI~700     & experimental & 12.8 & 20.2 & 29.3 & 37.3 & 48.0 \\
              & predicted    & 13.7 & 23.8 & 37.8 & 53.3 & 67.2\\
              &    &  &   & &  &  \\
    Costas~21  & experimental & 15.8 & 26.4 & 60.0 & 159.2 & 290.5 \\
              & predicted    & 16.0 & 32.0 & 64.0 & 128.0 & 256.0 \\
    \hline%
  \end{tabular}
  \caption{Comparison: experimental and predicted speedups}
  \label{tab:comparison}
\end{table}

It is worth noticing that our model approximates the behaviors
of experimental results very closely, as shown by the predicted speed-ups
matching closely the real ones. Moreover we can see that on the three
benchmark programs, we needed to use three different types
of distribution (exponential, shifted exponential and lognormal),
in order to approximate the experimental data most closely.
This shows that our model is quite general
and can accommodate different types of parallel behaviors.

\begin{figure}[t]
  \begin{center}
    \includegraphics[width=0.4\textwidth]{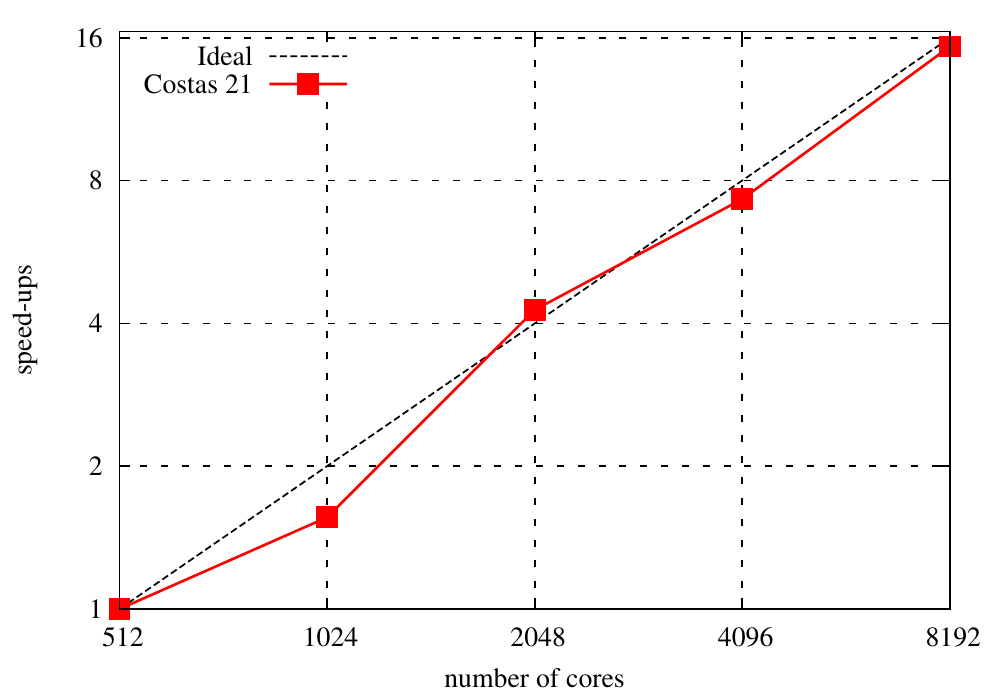}
    \caption{Speed-ups for Costas 21 up to 8,192 cores}
  \end{center}
  \label{fig:speedup-jugene}
\end{figure}

A quite interesting behavior is exhibited by the Costas 21 problem. Our
model predicts a linear speedup, up to 10,000 cores and beyond,
and the experimental data gathered for this paper confirms this linear
speed-up up to 256 cores. Would it scale up with a larger number of cores?
Indeed such an experiment has been done up to 8,192 cores on
the JUGENE  IBM Bluegene/P  at the J\"ulich Supercomputing Center in Germany
(with a total 294,912 cores),
and reported in~\cite{pco12}, of which Figure~\ref{fig:speedup-jugene} is
adapted. We can see that the speed-up is indeed linear up to 8,192 cores,
thus showing the adequation of the prediction model with the real data.

Finally, let us note that our method exhibits an interesting phenomenon.
For the three problems considered,  the  probability  of returning  a  solution  in
\emph{no} iterations is non-null: since they start by a uniform random
draw  on the  search  space, there  is  a very  small,  but not  null,
probability  that  this  random  initialization directly  returns  the
solution. Hence, in theory, $x_0=0$ and the speed-up should be linear,
with  an   infinite  limit   when  the  number   of  cores   tends  to
infinity. Intuitively,  if the number  of cores tends to  infinity, at
some point it will be large compared to the size of the search space
and one of the cores is likely to immediately find the solution.

Yet, in practice, observations  shows that the experimental curves may
be better approximated by a shifted exponential with $x_0>0$, as it is
the case for AI~700.
With an  exponential distribution,  this leads to  non-linear speed-up
with a finite limit.  Indeed,  the experimental speed-up for AI~700 is
far from linear.  On the contrary, Costas~21 has a linear speed-up due
to its  $x_0 << 1/\lambda$,  which makes the statistical  test succeed
for $x_0 \simeq 0$. Firstly, this suggests that the comparison between
$x_0$ and $1/\lambda$ on a number of observations is a key element for
the parallel behavior.  It also  means that the number of observations
needed  to properly approximate  the sequential  distribution probably
depends on  the problem.

\section{Conclusion}

We have proposed a theoretical  model for predicting and analyzing the
speed-ups  of   Las  Vegas   algorithms.  It is worth noticing that our
model mimics  the behaviors of the experimental results very closely,
as shown by the predicted speedups matching closely the real ones.
Our practical experiments consisted in testing the accuracy of the model
with respect to three instances of a local search algorithm 
for combinatorial optimization problems.
We showed that the parallel speed-ups predicted by our statistical
model are  accurate, matching the actual speed-ups very well up to
64 parallel cores and then with a deviation of about 10\%, 15\% or
30\% (depending on the benchmark problem) up to 256 cores.

However, one limitation of our approach is that, in practice, we need to
be able to compute the  expectation of  the minimum distribution.
Nevertheless, apart from  the exponential  distribution for  which this
computation is easy, recent  results  in  the  field  of order  statistics
gives  explicit formulas for  a number  of useful distributions:
gaussian, lognormal, gamma,  beta.
This provides a wide  range of  tools to analyze  different  behaviors.
In this paper we validated  our approach on classical  combinatorial optimization
and CSP benchmarks, but further  research  will consider a larger class of
problems and algorithms, such as SAT solvers and other randomized
algorithms (e.g. quick sort).

Another interesting extension of this work would be to devise  a method for
predicting the  speed-up from scratch, that is,  without any knowledge
on the algorithm  distribution.  Preliminary observation suggests that,
given  a  problem  and  an  algorithm,  the  general   shape  of  the
distribution is  the same when the  size of the  instances varies.
For example, the  different instances of \ai  that we tested  all admit a
shifted exponential distribution.  If this property is valid on a wide
range  of  problems/algorithms,  then  we  can develop  a  method  for
predicting the speed-up for large instances by learning the distribution shape
on small  instances (which are easier to solve), and then estimating
the parallel speed-up for larger instances with our model.


\begin{thebibliography}{40}
\providecommand{\natexlab}[1]{#1}
\providecommand{\url}[1]{\texttt{#1}}
\expandafter\ifx\csname urlstyle\endcsname\relax
  \providecommand{\doi}[1]{doi: #1}\else
  \providecommand{\doi}{doi: \begingroup \urlstyle{rm}\Url}\fi

\bibitem[Aarts and Lenstra(1997)]{aarts:1997:lsco}
E.~Aarts and J.~K. Lenstra, editors.
\newblock \emph{Local Search in Combinatorial Optimization}.
\newblock John Wiley and Sons, Chichester, UK, 1997.

\bibitem[Aiex et~al.(2007)Aiex, Resende, and Ribeiro]{tttplots2007}
R.~Aiex, M.~Resende, and C.~Ribeiro.
\newblock {TTT} plots: a perl program to create time-to-target plots.
\newblock \emph{Optimization Letters}, 1:\penalty0 355--366, 2007.
\newblock ISSN 1862-4472.

\bibitem[Aiex et~al.(2002)Aiex, Resende, and
  Ribeiro]{DBLP:journals/heuristics/AiexRR02}
R.~M. Aiex, M.~G.~C. Resende, and C.~C. Ribeiro.
\newblock Probability distribution of solution time in grasp: An experimental
  investigation.
\newblock \emph{Journal of Heuristics}, 8\penalty0 (3):\penalty0 343--373,
  2002.

\bibitem[Alba(2004)]{JHeuristics2004}
E.~Alba.
\newblock Special issue on new advances on parallel meta-heuristics for complex
  problems.
\newblock \emph{Journal of Heuristics}, 10\penalty0 (3):\penalty0 239--380,
  2004.

\bibitem[Babai(1979)]{Babai79}
L.~Babai.
\newblock Monte-carlo algorithms in graph isomorphism testing.
\newblock Research Report D.M.S. No. 79-10, Universit\'e de Montr\'eal, 1979.

\bibitem[Beard et~al.(2007)Beard, Russo, Erickson, Monteleone, and
  Wright]{4285351}
J.~Beard, J.~Russo, K.~Erickson, M.~Monteleone, and M.~Wright.
\newblock Costas array generation and search methodology.
\newblock \emph{Aerospace and Electronic Systems, IEEE Transactions on},
  43\penalty0 (2):\penalty0 522 --538, april 2007.
\newblock ISSN 0018-9251.
\newblock \doi{10.1109/TAES.2007.4285351}.

\bibitem[Bessiere(2006)]{Bessiere06}
C.~Bessiere.
\newblock Constraint propagation.
\newblock In F.~Rossi, P.~van Beek, and T.~Walsh, editors, \emph{Handbook of
  Constraint Programming}, pages 29--83. Elsevier, 2006.

\bibitem[Bolze and al.(2006)]{Grid5000}
R.~Bolze and al.
\newblock Grid'5000: A large scale and highly reconfigurable experimental grid
  testbed.
\newblock \emph{Int. J. High Perform. Comput. Appl.}, 20\penalty0 (4):\penalty0
  481--494, 2006.

\bibitem[Caniou et~al.(2012)Caniou, Diaz, Richoux, Codognet, and
  Abreu]{ppopp12}
Y.~Caniou, D.~Diaz, F.~Richoux, P.~Codognet, and S.~Abreu.
\newblock Performance analysis of parallel constraint-based local search.
\newblock In \emph{{PPoPP 2012}, 17th ACM SIGPLAN Symposium on Principles and
  Practice of Parallel Programming}, New Orleans, LA, USA, 2012. ACM Press.
\newblock poster paper.

\bibitem[Codognet and Diaz(2001)]{DBLP:conf/saga/CodognetD01}
P.~Codognet and D.~Diaz.
\newblock Yet another local search method for constraint solving.
\newblock In \emph{proceedings of SAGA'01}, pages 73--90. Springer Verlag,
  2001.

\bibitem[Codognet and Diaz(2003)]{mic/CodognetD03}
P.~Codognet and D.~Diaz.
\newblock An efficient library for solving {CSP} with local search.
\newblock In T.~Ibaraki, editor, \emph{MIC'03, 5th International Conference on
  Metaheuristics}, 2003.

\bibitem[Costas(1984)]{Cos-84}
J.~Costas.
\newblock A study of detection waveforms having nearly ideal range-doppler
  ambiguity properties.
\newblock \emph{Proceedings of the IEEE}, 72\penalty0 (8):\penalty0 996--1009,
  1984.

\bibitem[Crainic and Toulouse(2002)]{JHeuristics2002}
T.~Crainic and M.~Toulouse.
\newblock Special issue on parallel meta-heuristics.
\newblock \emph{Journal of Heuristics}, 8\penalty0 (3):\penalty0 247--388,
  2002.

\bibitem[Crainic et~al.(2004)Crainic, Gendreau, Hansen, and
  Mladenovic]{DBLP:journals/heuristics/CrainicGHM04}
T.~G. Crainic, M.~Gendreau, P.~Hansen, and N.~Mladenovic.
\newblock Cooperative parallel variable neighborhood search for the {\it
  }-median.
\newblock \emph{Journal of Heuristics}, 10\penalty0 (3):\penalty0 293--314,
  2004.

\bibitem[David and Nagaraja(2003)]{david2003order}
H.~David and H.~Nagaraja.
\newblock \emph{Order Statistics}.
\newblock Wiley series in probability and mathematical statistics. Probability
  and mathematical statistics. John Wiley, 2003.
\newblock ISBN 9780471389262.

\bibitem[Diaz et~al.(2012)Diaz, Richoux, Caniou, Codognet, and Abreu]{pco12}
D.~Diaz, F.~Richoux, Y.~Caniou, P.~Codognet, and S.~Abreu.
\newblock Parallel local search for the costas array problem.
\newblock In \emph{IEEE Workshop on new trends in Parallel Computing and
  Optimization (PC012), in conjunction with IPDPS 2012}, Shanghai, China, May
  2012. IEEE Press.

\bibitem[Drakakis(2006)]{Dra-06}
K.~Drakakis.
\newblock A review of costas arrays.
\newblock \emph{Journal of Applied Mathematics}, 2006:\penalty0 1--32, 2006.

\bibitem[Drakakis et~al.(2011{\natexlab{a}})Drakakis, Iorio, and Rickard]{n28}
K.~Drakakis, F.~Iorio, and S.~Rickard.
\newblock The enumeration of costas arrays of order 28 and its consequences.
\newblock \emph{Advances in Mathematics of Communications}, 5\penalty0
  (1):\penalty0 69--86, 2011{\natexlab{a}}.

\bibitem[Drakakis et~al.(2011{\natexlab{b}})Drakakis, Iorio, Rickard, and
  Walsh]{n29}
K.~Drakakis, F.~Iorio, S.~Rickard, and J.~Walsh.
\newblock Results of the enumeration of costas arrays of order 29.
\newblock \emph{Advances in Mathematics of Communications}, 5\penalty0
  (3):\penalty0 547--553, 2011{\natexlab{b}}.

\bibitem[Eadie(1971)]{eadie1971statistical}
W.~Eadie.
\newblock \emph{Statistical methods in experimental physics}.
\newblock North-Holland Pub. Co., 1971.

\bibitem[Galinier and Hao(2000)]{Hao00}
P.~Galinier and J.-K. Hao.
\newblock A general approach for constraint solving by local search.
\newblock In \emph{2nd workshop CP-AI-OR'00}, Paderborn, Germany, 2000.

\bibitem[Gent and Walsh(1999)]{CSPLIB}
I.~P. Gent and T.~Walsh.
\newblock {CSPLIB}: A benchmark library for constraints.
\newblock In \emph{proceedings of CP'99}, pages 480--481. Springer Verlag,
  1999.

\bibitem[Golomb(1984)]{Gol-84}
S.~Golomb.
\newblock Algebraic constructions for {C}ostas arrays.
\newblock \emph{Journal Of Combinatorial Theory Series A}, 37\penalty0
  (1):\penalty0 13--21, 1984.

\bibitem[Golomb and Taylor(1984)]{GT-84}
S.~Golomb and H.~Taylor.
\newblock Constructions and properties of {C}ostas arrays.
\newblock \emph{Proceedings of the IEEE}, 72\penalty0 (9):\penalty0 1143--1163,
  1984.

\bibitem[Gomes and Selman(2001)]{portfolio01}
C.~P. Gomes and B.~Selman.
\newblock Algorithm portfolios.
\newblock \emph{Artificial Intelligence}, 126\penalty0 (1-2):\penalty0 43--62,
  2001.

\bibitem[Gonzalez(2007)]{handbook-approx}
T.~Gonzalez, editor.
\newblock \emph{Handbook of Approximation Algorithms and Metaheuristics}.
\newblock Chapman and Hall / CRC, 2007.

\bibitem[Hentenryck(1989)]{PVH89}
P.~V. Hentenryck.
\newblock \emph{Constraint Satisfaction in Logic Programming}.
\newblock The MIT Press, 1989.

\bibitem[Hentenryck and Michel(2005)]{comet05}
P.~V. Hentenryck and L.~Michel.
\newblock \emph{Constraint-Based Local Search}.
\newblock The MIT Press, 2005.

\bibitem[Hoos and St\"utze(2005)]{Hoos-book}
H.~Hoos and T.~St\"utze.
\newblock \emph{Stochastic Local Search: Foundations and Applications}.
\newblock Morgan Kaufmann, 2005.

\bibitem[Ibaraki et~al.(2005)Ibaraki, Nonobe, and Yagiura]{metaheuristics}
T.~Ibaraki, K.~Nonobe, and M.~Yagiura, editors.
\newblock \emph{Metaheuristics: Progress as Real Problem Solvers}.
\newblock Springer Verlag, 2005.

\bibitem[Lin(1965)]{Lin65}
S.~Lin.
\newblock Computer solutions of the traveling salesman problem.
\newblock \emph{Bell System Technical Journal}, 44:\penalty0 2245--2269, 1965.

\bibitem[Minton et~al.(1992)Minton, Johnston, Philips, and Laird]{min-conflict}
S.~Minton, M.~D. Johnston, A.~B. Philips, and P.~Laird.
\newblock Minimizing conflicts: A heuristic repair method for constraint
  satisfaction and scheduling problems.
\newblock \emph{Artificial Intelligence}, 58\penalty0 (1-3):\penalty0 161--205,
  1992.

\bibitem[Nadarajah(2008)]{Nadarajah2008}
S.~Nadarajah.
\newblock Explicit expressions for moments of order statistics.
\newblock \emph{Statistics \& Probability Letters}, 78\penalty0 (2):\penalty0
  196--205, Feb. 2008.

\bibitem[Pardalos et~al.(1995)Pardalos, Pitsoulis, Mavridou, and
  Resende]{DBLP:conf/irregular/PardalosPMR95}
P.~M. Pardalos, L.~S. Pitsoulis, T.~D. Mavridou, and M.~G.~C. Resende.
\newblock Parallel search for combinatorial optimization: Genetic algorithms,
  simulated annealing, tabu search and {GRASP}.
\newblock In \emph{proceedings of IRREGULAR}, pages 317--331, 1995.

\bibitem[Russo et~al.(2010)Russo, Erickson, and
  Beard]{DBLP:conf/ciss/RussoEB10}
J.~C. Russo, K.~G. Erickson, and J.~K. Beard.
\newblock Costas array search technique that maximizes backtrack and symmetry
  exploitation.
\newblock In \emph{CISS}, pages 1--8, 2010.

\bibitem[Truchet(2004)]{Truchet04}
C.~Truchet.
\newblock \emph{Constraints, Local Search and Computer-Aided Music
  Composition}.
\newblock PhD thesis, University of Paris 7, 2004.

\bibitem[Truchet and Codognet(2004)]{DBLP:journals/soco/TruchetC04}
C.~Truchet and P.~Codognet.
\newblock Musical constraint satisfaction problems solved with adaptive search.
\newblock \emph{Soft Comput.}, 8\penalty0 (9):\penalty0 633--640, 2004.

\bibitem[Van~Luong et~al.(2010)Van~Luong, Melab, and Talbi]{GPU2010}
T.~Van~Luong, N.~Melab, and E.-G. Talbi.
\newblock Local search algorithms on graphics processing units.
\newblock In \emph{Evolutionary Computation in Combinatorial Optimization},
  pages 264--275. LNCS 6022, Springer Verlag, 2010.

\bibitem[Verhoeven and Aarts(1995)]{Aarts95}
M.~Verhoeven and E.~Aarts.
\newblock Parallel local search.
\newblock \emph{Journal of Heuristics}, 1\penalty0 (1):\penalty0 43--65, 1995.

\bibitem[Wolfram(2003)]{mathematica}
S.~Wolfram.
\newblock \emph{The {M}athematica Book, 5th edition}.
\newblock Wolfram Media, 2003.
\newblock URL \url{http://reference.wolfram.com}.

\end{thebibliography}


\end{document}